\documentclass[aps, prb, reprint, superscriptaddress]{revtex4-2}

\usepackage{graphicx,color,amssymb}
\usepackage{xcolor}
\usepackage{hyperref}

\newcommand{\eps}{\varepsilon}      

\begin{document}

\title{Effective Medium Perspective on Topological Transitions in Metamaterials}

\author{Leon Shaposhnikov}
\thanks{These two authors contributed equally}
\affiliation{School of Physics and Engineering, ITMO University, Saint Petersburg, Russia}

\author{Denis Sakhno}
\thanks{These two authors contributed equally}
\affiliation{School of Physics and Engineering, ITMO University, Saint Petersburg, Russia}

\author{Daniel A. Bobylev}
\affiliation{School of Physics and Engineering, ITMO University, Saint Petersburg, Russia}

\author{Maxim A. Gorlach}
\email{m.gorlach@metalab.ifmo.ru}
\affiliation{School of Physics and Engineering, ITMO University, Saint Petersburg, Russia}

\begin{abstract}
Many properties of photonic structures rely on band topology characterized by the integer invariants that can change during the topological transitions and give rise to the disorder-robust topological edge, corner, or interface states. Typically the periods of such structures are comparable to the wavelength. However, in many cases, the unit cell becomes deeply subwavelength and hence the entire metamaterial can be described in terms of the effective material parameters. Here, focusing on subwavelength topological metamaterials, we identify the behavior of permittivity and permeability accompanying the topological transition on the example of the two structures possessing $D_6$ symmetry.
\end{abstract}

\maketitle

\section{Introduction}\label{sec:Intro}

The concept of band topology has recently become a powerful tool in diverse areas of physics ranging from condensed matter~\cite{Xiao2010,Hasan-Kane}, photonics~\cite{Lu2014,Ozawa2019} and polaritonics~\cite{Karzig,Klembt2018} to acoustics~\cite{Yang2015} and mechanics~\cite{Huber2016}. The topological properties are captured by the integer invariants which are robust against local perturbations of the structure and subject to change only during the topological transitions accompanied by closing and reopening of bandgaps in the spectrum. In turn, these conserved quantities give rise to the topological states featuring a remarkable robustness to disorder and imperfections.

In photonics, such topological structures have been realized throughout the entire electromagnetic spectrum, from microwaves to the visible~\cite{Ozawa2019}. In most of the cases, their periods are comparable to wavelength, and the topological properties are retrieved from the behavior of the Bloch functions at the different points of the Brillouin zone in analogy to condensed matter systems~\cite{Benalcazar2019}. An opposite situation is the effective medium limit when the Brillouin zone is ill-defined and the topology should be tested via different techniques~\cite{Silv2015,Silveirinha2016,Silv2016,Hanson2022}. Proper assessment of this limit requires additional assumptions such as the behavior of the permittivity at large wave numbers~\cite{Silveirinha2016}. However, the connection between these two limits is currently lacking and the possibility to treat subwavelength topological metamaterials from the effective medium perspective remains questionable in many aspects.

\begin{figure}[b]
	\begin{minipage}{0.9\linewidth}
		\center{\includegraphics[width=1\textwidth]{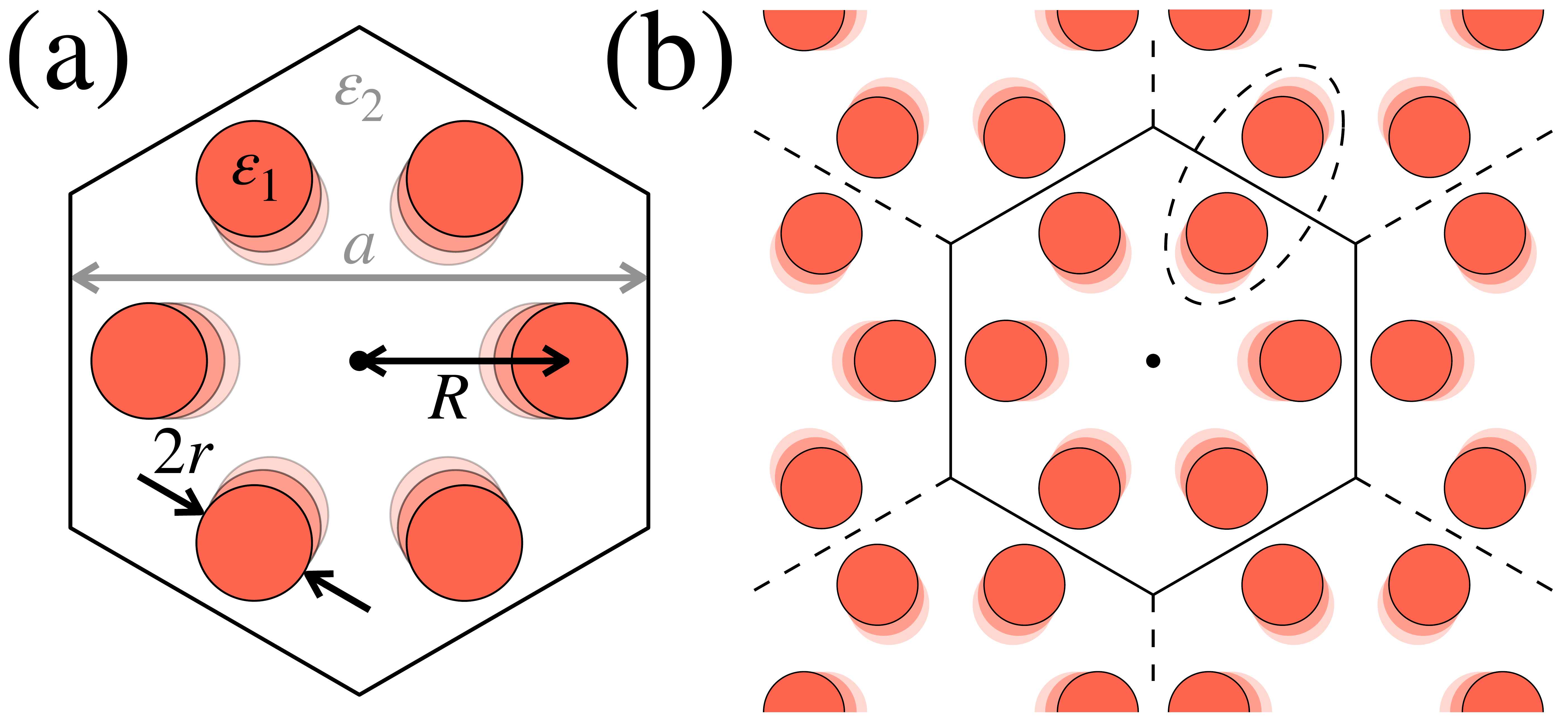}}
	\end{minipage}
	\caption{Unit cell (a) and geometry (b) of the topological structure based on the array of identical dielectric rods placed in the sites of breathing honeycomb lattice. The topological properties are controlled by the $a/R$ ratio.}
	\label{fig:3c-geom}
\end{figure}

In this Article, we aim to fill in this gap by studying the paradigmatic topological metamaterial based on the breathing honeycomb lattice~\cite{Wu}, Fig.~\ref{fig:3c-geom}. Due to its simplicity, this topological  structure has been realized in numerous experiments from microwaves~\cite{Yves2017,Li2018,Yang2018} to optics~\cite{Barik2018,Noh2018,Gorlach2018,Smirnova2019,Kuipers2020}. Importantly, its topological properties are determined mainly by the vicinity of the $\Gamma$ point in the Brillouin zone and hence are dominated by the low-$k$ modes. As a result, the period of this structure can be made subwavelength and the entire topological metamaterial approaches the effective medium limit.

Exploiting this, in Sec.~\ref{sec:3C} we retrieve the effective permittivity and permeability of the structure via Nicolson-Ross-Weir (NRW)  technique~\cite{Nicolson,Weir} and trace their evolution with the change of the lattice parameters. After establishing the connection between the change of material parameters and the topological transition, we use this observation and reveal another topological metamaterial in Sec.~\ref{sec:4B}. For consistency, we also support this identification by calculating the invariants via the conventional procedure. Finally, in Sec.~\ref{sec:Discussion} we discuss the consequences and advantages of our results highlighting further possible directions of research.

\section{Effective material parameters for breathing honeycomb lattice}\label{sec:3C}

We start by revisiting the properties of breathing honeycomb lattice with the unit cell depicted in Fig.~\ref{fig:3c-geom}(a). The unit cell of such lattice [Fig.~\ref{fig:3c-geom}(b)] consists of six identical dielectric rods 
of radii $r$ arranged in a $C_6$-symmetric hexagon cluster, which is defined by the lattice constant, $a$, and the distance between the centers of the unit cell and a single rod, $R$. We assume that the lattice constant $a$ and the radii of the rods $r$ are fixed, while the size of the hexagon $R$ can be tuned. As a specific example, we consider the parameters $r = 0.5$~cm, $a = 5$~cm, $\eps_1 = 81$, $\eps_2 = 1$ that correspond to the water-based metamaterial in the microwave range.

\begin{figure}[t]
	\begin{minipage}{0.9\linewidth}
		\center{\includegraphics[width=1\textwidth]{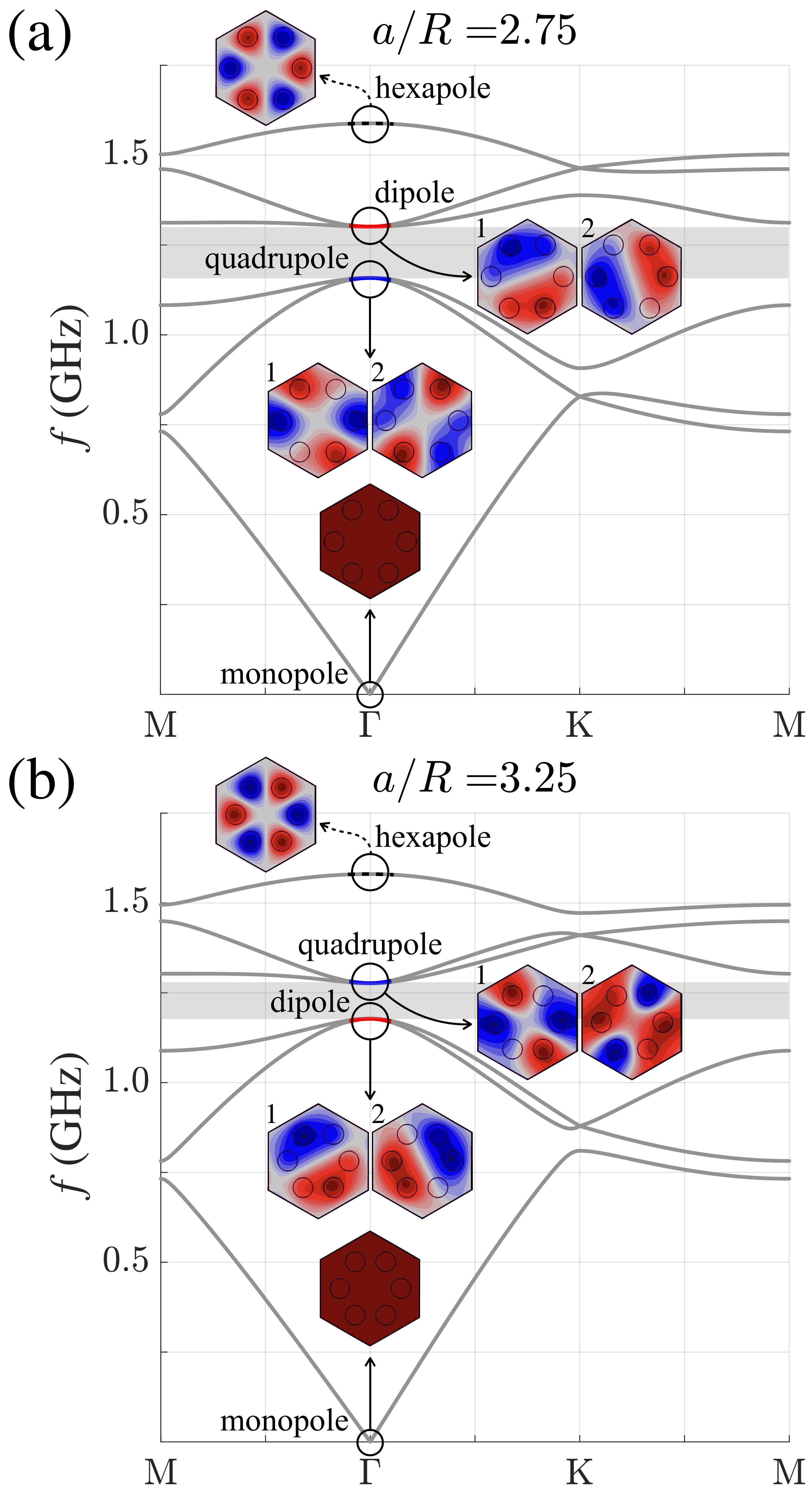}}
	\end{minipage}
	\caption{Band structure for the breathing honeycomb lattice with geometric parameters (a) $a/R=2.75$ (expanded)  and (b) $a/R=3.25$ (shrunken) calculated with Comsol Multiphysics package. Complete photonic bandgap is shaded by gray. Insets show the characteristic distribution of electric field component $E_z$ within the unit cell for each of the bands. The calculations are performed for the structure with $a=5$~cm, $r=0.5$~cm, permittivity of the rods $\varepsilon_{1}=81$, permittivity of the host medium $\eps_{2}=1$.}
	\label{fig:3c-disp}
\end{figure}

In the case of $a/R = 3$ the structure reduces to a conventional graphene-like honeycomb lattice with two rods in the primitive cell. If the size of the unit cell is increased up to six cylinders [Fig.~\ref{fig:3c-geom}(a)], the Brillouin zone folds and the dispersion branches for the four modes  cross at the $\Gamma$ point forming the Dirac point. Due to the symmetry of the structure, the modes at $\Gamma$ point can be distinguished by their behavior relative to $C_6$ rotations. Specifically, the modes which acquire the phases $0$, $\pm \pi/3$, $\pm 2\pi/3$ and $\pi$ under $C_6$ rotation are referred below as monopole, dipole, quadrupole and hexapole modes. Note that the dipole and quadrupole modes are doubly degenerate.


If the ratio $a/R$ is different from 3, the primitive cell includes six rods and the degeneracy of the bands is partially lifted giving rise to the complete bandgap [Fig.~\ref{fig:3c-disp}]. The cases $a/R > 3$ and $a/R < 3$ are referred to as  shrunken and expanded honeycomb lattices, respectively. When the system experiences a smooth transition from the expanded [Fig.~\ref{fig:3c-disp}(a)] to shrunken   [Fig.~\ref{fig:3c-disp}(b)] geometry, the inversion of dipole and quadrupole bands occurs indicating the topological transition~\cite{Wu}.

The change of the bands topology can be rigorously confirmed by the direct evaluation of the topological indices~\cite{Benalcazar2019} for both structures. In $C_6$-symmetric case, the indices are defined as
\begin{equation}\label{eq:symmind}
   \chi^{(6)}=(\#M_1^{(2)}-\#\Gamma_1^{(2)},\#K_1^{(3)}-\#\Gamma_1^{(3)})\:, 
\end{equation}
where $\#\Pi_m^{(n)}$ denotes the number of the modes which acquire $\exp(2\pi i(m-1)/n)$ phase factor under $C_n$ rotation at the high-symmetry point $\Pi$. The results for the expanded and shrunken geometries read:
\begin{equation}\label{eq:ind-3C}
    \chi_{{\rm e}}^{(6)} = (-2,0)\:,\quad \chi_{{\rm s}}^{(6)} = (0,0).
\end{equation}

\begin{figure}[t]
	\begin{minipage}{0.9\linewidth}
		\center{\includegraphics[width=1\textwidth]{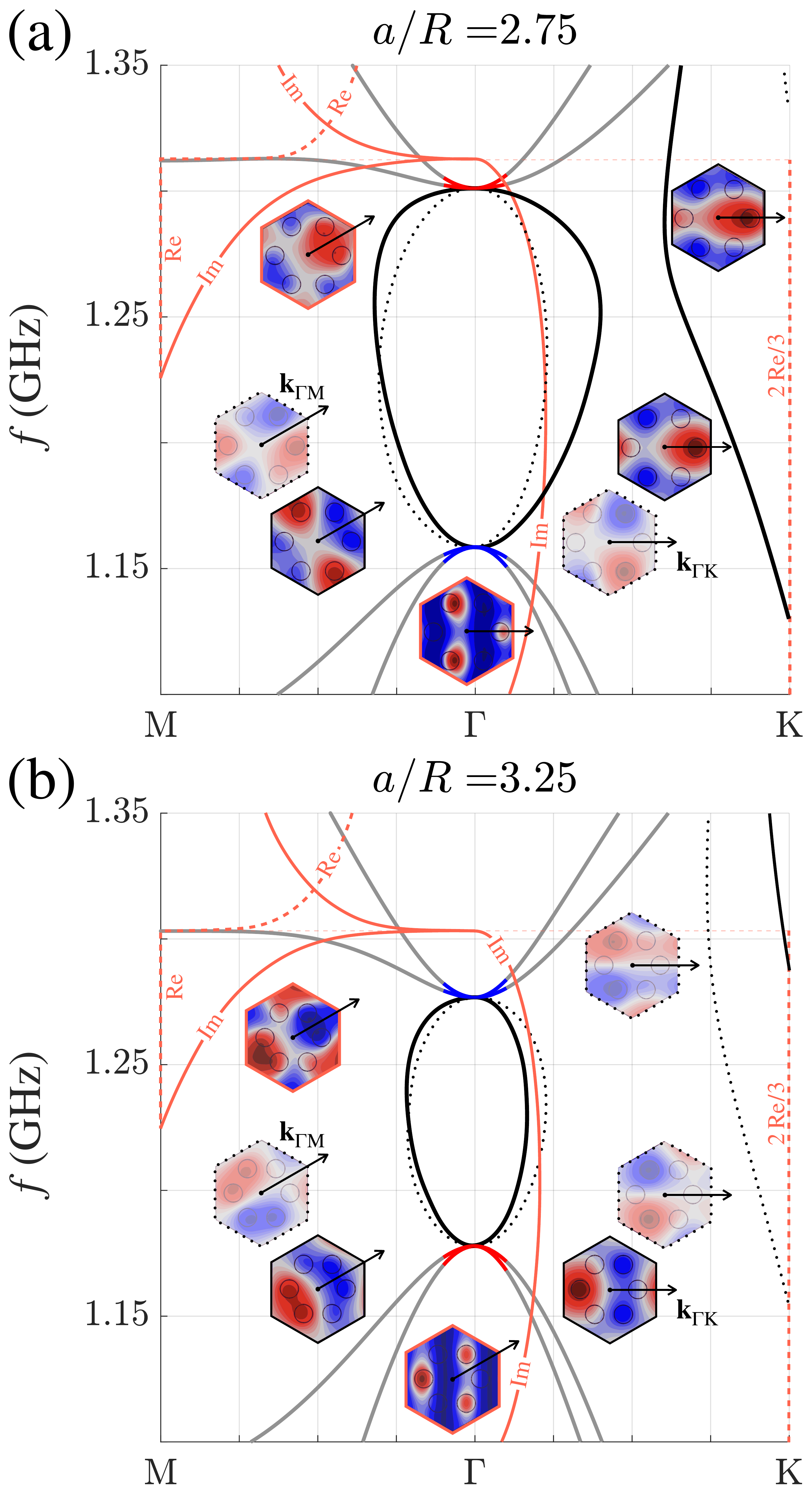}}
	\end{minipage}
	\caption{Dispersion diagrams for the evanescent waves in the band gap of breathing honeycomb lattice with (a) $a/R=2.75$ (expanded) and (b) $a/R=3.25$ (shrunken). Solid gray lines show modes with the real wave number. Eigenmodes with purely imaginary wave numbers $i\,k''$ are shown by black solid and dotted black lines depending on the fact whether the mode can be excited at normal incidence or not.  Orange solid and dashed lines show imaginary and real part of the wave number for the complex modes ($k'+i\,k''$). Insets show the distribution of $E_z$ component of electric field within the unit cell. Bright and dim insets correspond to the modes that can or cannot be excited at normal incidence.}
	\label{fig:3c-disp-weak}
\end{figure}

However, the dispersion diagrams in Fig.~\ref{fig:3c-disp} miss the evanescent modes defining the response of the structure at frequencies within the bulk bandgap. To complement the description, we analyze complex band diagrams obtained via full-wave numerical eigenmodes simulations. In contrast to the conventional eigenvalue problem for a periodic structure solved via common $f(\mathbf{k})$ formulation in the frequency domain, here we consider Bloch wave number in fixed $\Gamma$M or $\Gamma$K directions as an eigenvalue solved for, whereas the frequency $f$ is a parameter swept through the entire bandgap. By solving such $k(f)$ problem via \textit{Weak Form PDE} interface of Comsol Multiphysics software package~\cite{davancco2007complex, fietz2011complex}, we recover the set of modes with complex wave numbers at a given frequency. The obtained results are presented in Fig.~\ref{fig:3c-disp-weak} featuring a complicated interplay of the different types of modes. Three main types of the modes are distinguished:
%
%
%
%
\begin{enumerate}
    \item Modes with a purely real wave vector ($\mathbf{k}=k'\mathbf{\hat k}$, where $\mathbf{\hat k}$ is the unit vector in the chosen direction and $k'\in\mathbb{R}$) are depicted by solid gray lines. These are the conventional bulk modes shown in Fig.~\ref{fig:3c-disp}.
    \item Modes with a purely imaginary wave vector ($\mathbf{k}=i\,k''\mathbf{\hat k}$, where $k''\in\mathbb{R}$) are depicted by
    \begin{enumerate}
        \item black solid lines if the field distribution within the unit cell  is symmetric with respect to the $\mathbf{\hat k}$ direction, i.e. the mode can be excited by the impinging plane wave at normal incidence (see insets in Fig. \ref{fig:3c-disp-weak}).
        \item black dotted lines if the field distribution is anti-symmetric with respect to the $\mathbf{\hat k}$ direction, i.e. the mode cannot be excited by the impinging wave at normal incidence.
    \end{enumerate}
    \item Modes with a complex wave vector $\mathbf{k}=(k'+ik'')\mathbf{\hat k}$, where $k',k''\in\mathbb{R}$ and $k',k''\neq0$. Their imaginary and real parts are shown by solid and dashed lines, respectively and signed as $\mathrm{Im}$ or $\mathrm{Re}$. The orange color stands for the modes that can be excited.
\end{enumerate}


\begin{figure}[t]
	\begin{minipage}{0.9\linewidth}
		\center{\includegraphics[width=1\textwidth]{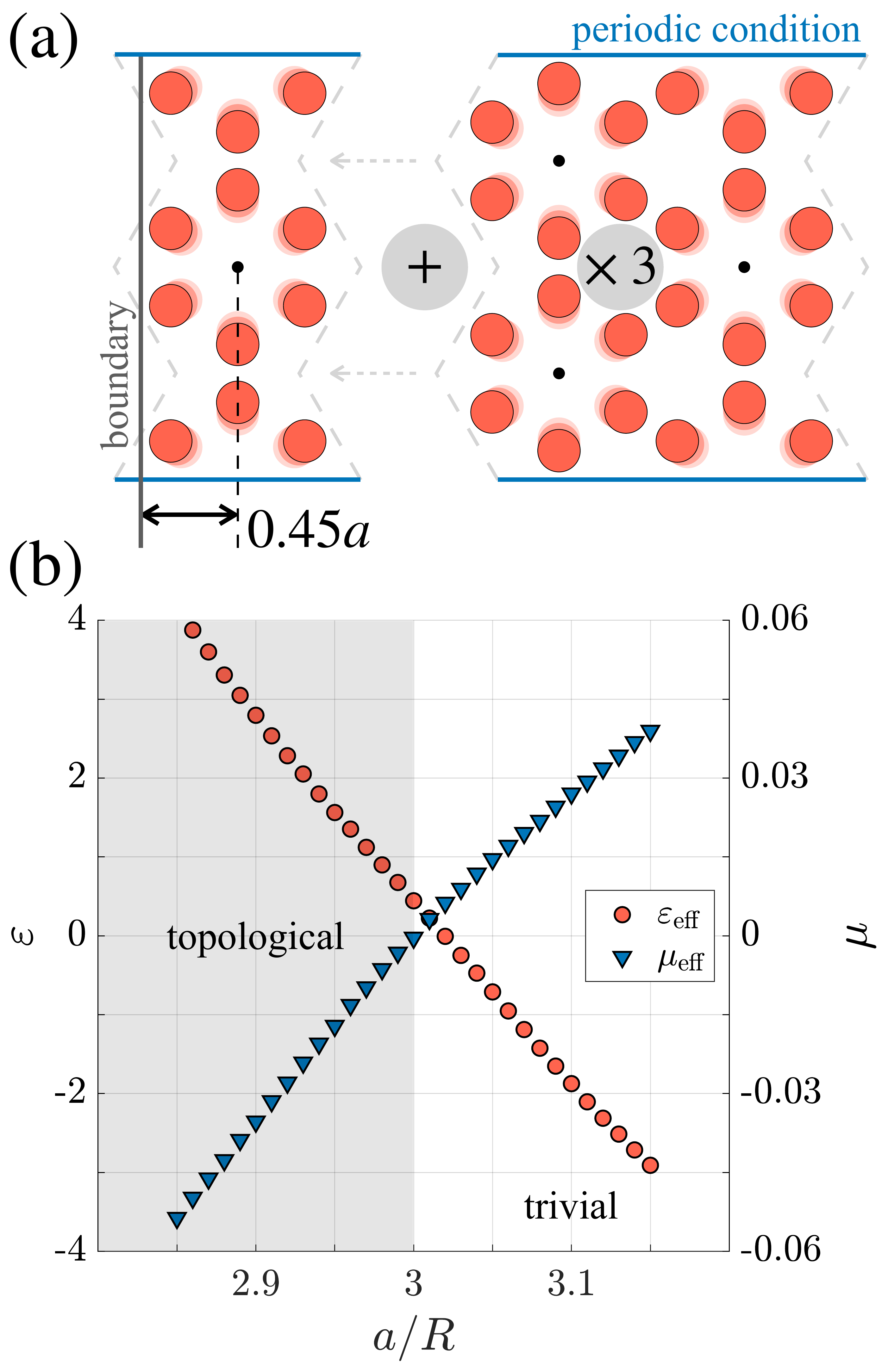}}
	\end{minipage}
	\caption{(a) Geometry of the slab used to calculate the reflection and transmission coefficients ($S_{11}$ and $S_{12}$ parameters). The normal to the slab corresponds to $\Gamma$M direction. The choice of the metamaterial boundary is shown by the black solid line. The distance from the boundary to the center of the first unit cell is equal to $0.45\,a$. Periodic boundary conditions along the blue lines are applied.
    (b) The results of the retrieval procedure inside the complete band gap at frequency $f=1.225$~GHz. Topological and trivial regimes correspond to $\varepsilon>0$, $\mu<0$ and $\varepsilon<0$, $\mu>0$, respectively.}
	\label{fig:3c-extraction}
\end{figure}

If the frequency of the incident wave $f$ is close to the center-of-bandgap frequency $1.215$~GHz, and the lattice constant is chosen to be $a=5$~cm, the wavelength-to-period ratio is around 5, i.e. the unit cell is subwavelength. Thus, one may attempt the description of the structure in terms of frequency-dependent effective permittivity and permeability. However, the effective medium picture is applicable provided the structure supports a single dominant mode with the largest attenuation length (smallest $k''$) with the rest of the modes decaying much more rapidly.
%
%
Inspecting the dispersion diagram Fig.~\ref{fig:3c-disp-weak} and ignoring the modes not excited at normal incidence, we observe that these requirements are fulfilled for the $\Gamma$M direction in the lattice, while $\Gamma$K direction exhibits two- or three-mode regime hindering the effective medium approach.

Accordingly, we choose such termination of the metamaterial that the normal to the boundary corresponds to the $\Gamma$M crystallographic direction [Fig.~\ref{fig:3c-extraction}(a)]. Replacing the metamaterial by the homogeneous slab, we have to specify the thickness of this slab, i.e. choose the boundary of the structure. It should be noted that there is no general recipe for such a choice, since the boundary conditions for the averaged fields in the metamaterial can be quite different from those expected for the continuous medium~\cite{Gorlach2020}. To choose the boundary consistently, we apply two requirements. First, the extracted permittivity and permeability should be real since the structure under study is lossless. Second, the obtained propagation constant of the evanescent wave should agree with that computed via $k(f)$ approach [Fig.~\ref{fig:3c-disp-weak}(b)] as further discussed in Appendix~A. Based on that, we choose the boundary at the distance $0.45\,a$ from the first unit cell center, see Fig.~\ref{fig:3c-extraction}(a).  

Next, we apply Nicolson-Ross-Weir (NRW) method~\cite{Nicolson,Weir} to retrieve the effective material parameters of the sample at frequency $f=1.225$~GHz. As a result, we obtain the values of $\eps\,\mu$ and $\mu/\eps$, both having negative signs as expected within the bandgap. This indicates that the  permittivity and permeability have different signs. To distinguish which of the material parameters is positive and which is negative, we  simulate the reflection from thick metamaterial sample. Inspecting the phase of the reflection coefficient, we identify the signs of $\eps$ and $\mu$ as further discussed in Appendix~B.

The effective permittivity and permeability extracted in this way are presented in Fig.~\ref{fig:3c-extraction}(b) as a function of shrinking-expanding parameter $a/R$. If $a/R=3$, permittivity and permeability of the metamaterial are simultaneously zero which is a consequence of a vanishing bandgap and finite impedance. Topologically nontrivial expanded lattice [see Eq.~(\ref{eq:ind-3C})] possesses positive permittivity and negative permeability. In the trivial case of shrunken lattice, however, permittivity and permeability swap their signs and the permeability becomes positive. Note that these conclusions for TE-polarized modes are consistent with the calculations of Ref.~\cite{Silveirinha2016}. We anticipate that the observed behavior of material parameters is quite general, since in most of cases topological transitions are accompanied by closing and reopening of a bandgap.

\section{Detecting topological transitions via the retrieved material parameters}\label{sec:4B}

\begin{figure}[b]
	\begin{minipage}{0.9\linewidth}
		\center{\includegraphics[width=1\textwidth]{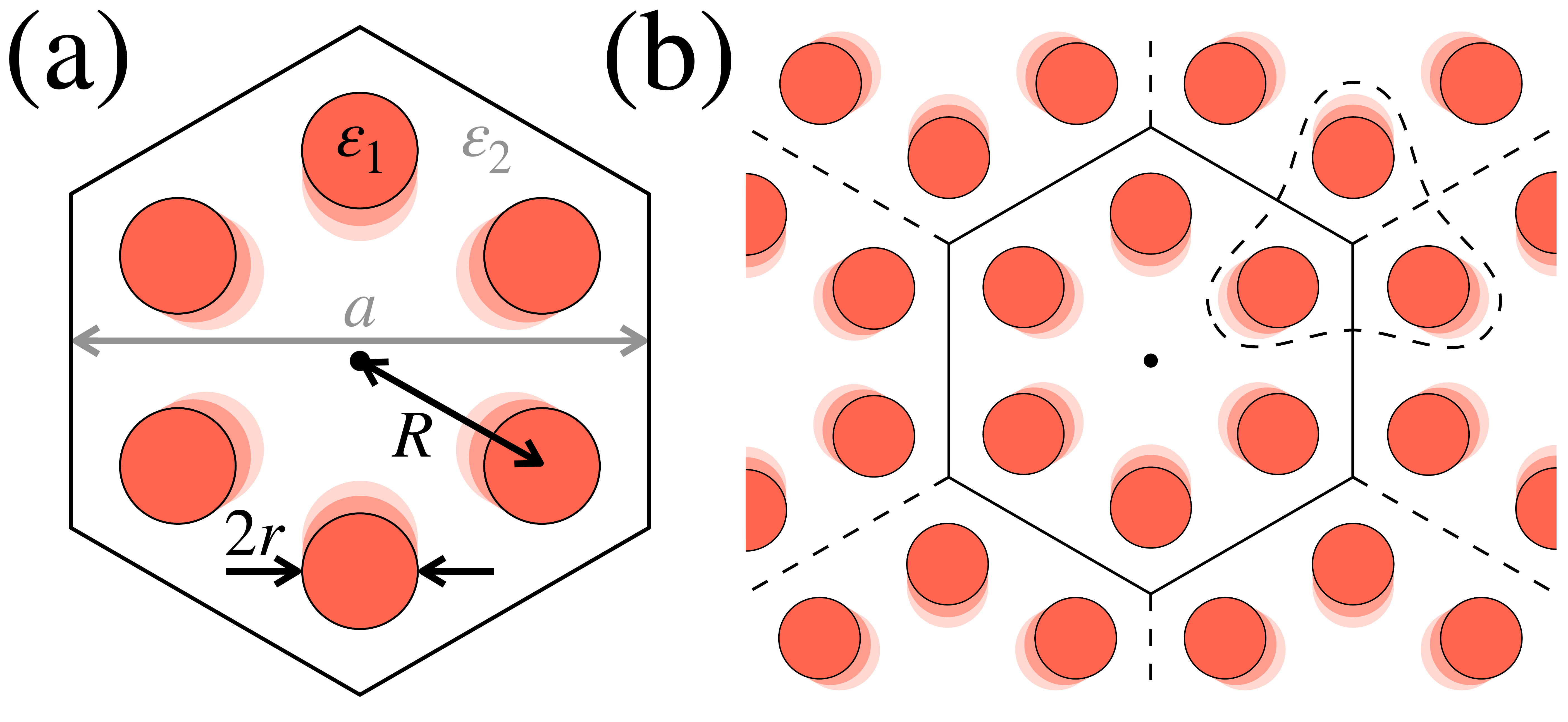}}
	\end{minipage}
	\caption{(a) Unit cell and (b) geometry of the modified breathing honeycomb lattice  based on the array of identical dielectric rods. The topological properties are controlled by the $a/R$ ratio.}
	\label{fig:4b-geom}
\end{figure}

The results of the previous section suggest that the topological transition can be accompanied by the simultaneous change of sign of the effective permittivity and permeability. Therefore, such a behavior can be used to diagnose topologically nontrivial metamaterials. To validate our approach, we consider a modified structure, Fig.~\ref{fig:4b-geom}, which differs from the original breathing honeycomb geometry (Fig.~\ref{fig:3c-geom}) by $\pi/6$ rotation of all six rods around the unit cell center.

\begin{figure}[t]
	\begin{minipage}{0.9\linewidth}
		\center{\includegraphics[width=1\textwidth]{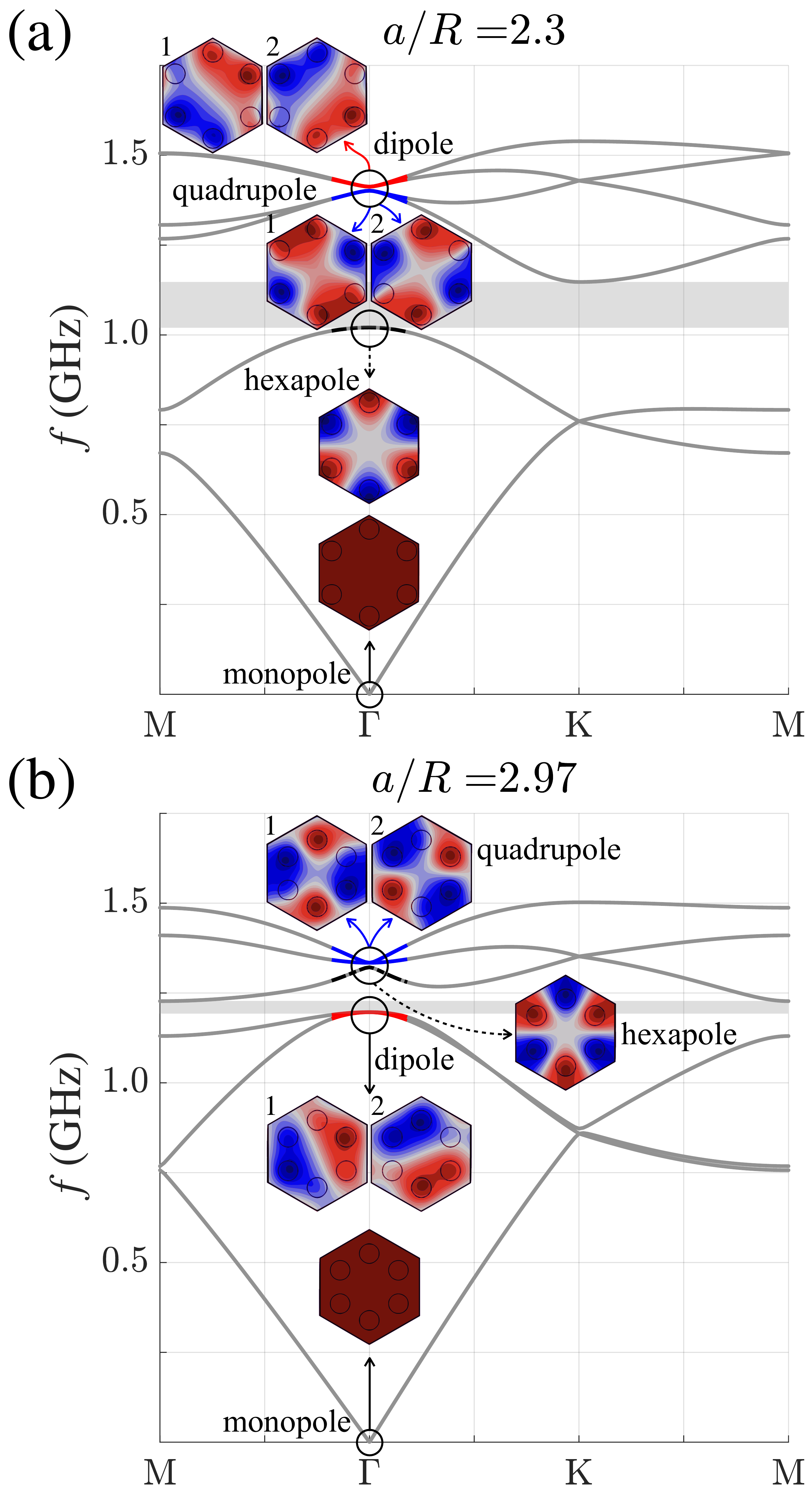}}
	\end{minipage}
	\caption{Dispersion diagrams for the modified breathing honeycomb lattice with the ratios of geometric parameters (a) $a/R=2.3$  and (b) $a/R=2.97$. Insets show the characteristic distribution of electric field $E_z$ component for each of the dispersion branches near $\Gamma$-point. All diagrams are plotted for the structure with $a=5$~cm, $r=0.5$~cm, permittivity of the rods $\varepsilon_{1}=81$ and host material $\eps_2=1$.}
	\label{fig:4b-disp}
\end{figure}

While such modification seems slight, it alters the dispersion of the modes and their order as it is evident from the calculated band diagrams [Fig.~\ref{fig:4b-disp}]. Specifically, the hexapole mode can now appear in-between dipole and quadrupole modes which determined previously  the topological nature of the model (Fig.~\ref{fig:3c-disp}). Therefore, chosen perturbation  affects the topological properties which are much less studied in this case~\cite{Wu2021}.

 
\begin{figure}[t]
	\begin{minipage}{0.9\linewidth}
		\center{\includegraphics[width=1\textwidth]{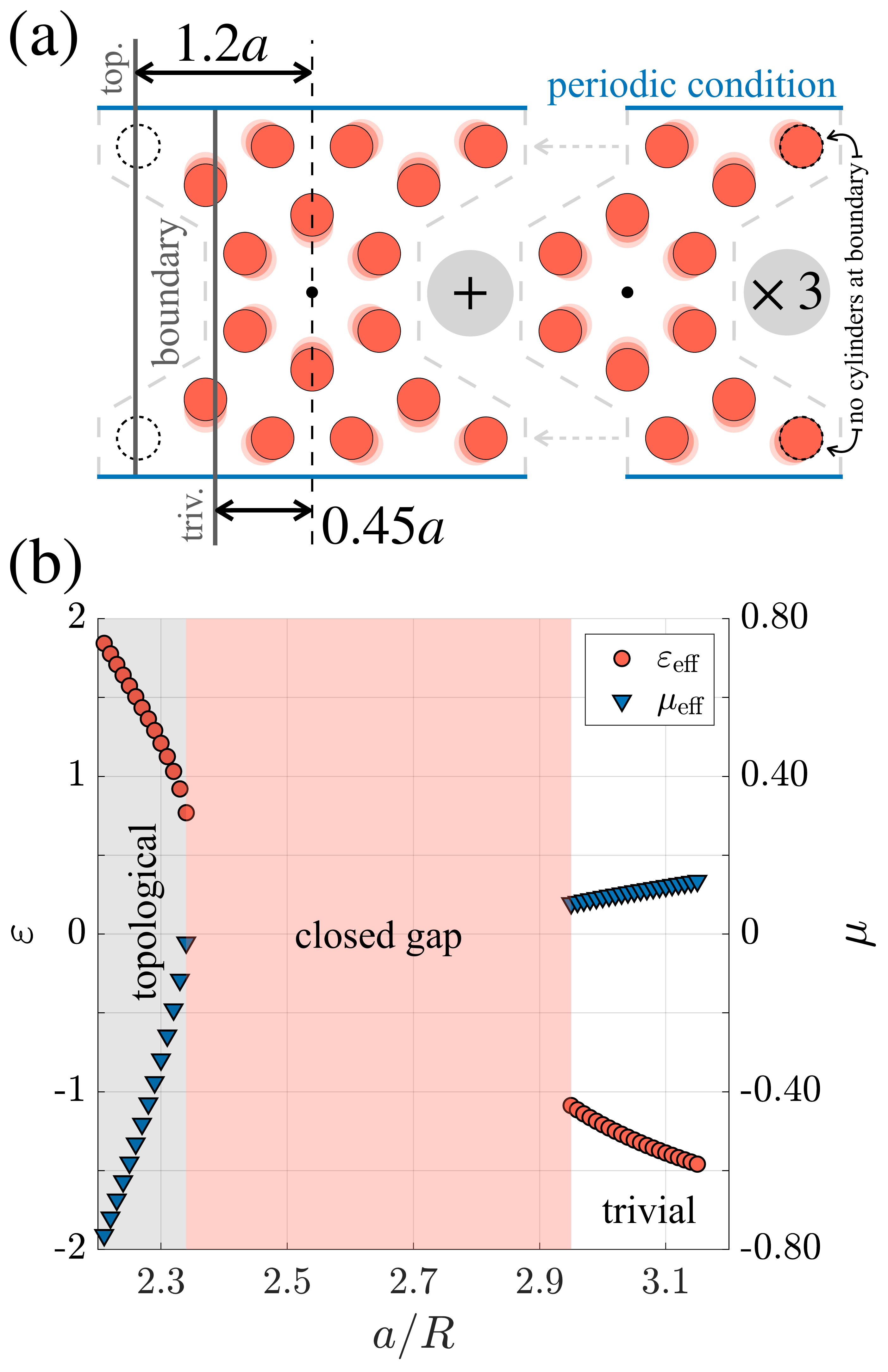}}
	\end{minipage}
	\caption{(a) Geometry of the slab used to calculate the reflection and transmission coefficients ($S_{11}$ and $S_{12}$ parameters). The normal to the slab corresponds to the $\Gamma$K direction. The choice of the metamaterial boundary for the frequencies $f_1=1.12$~GHz and $f_2=1.215$~GHz is shown by solid vertical lines: the distances to the reference plane shown by the black dashed line are $0.45a$ and $1.2a$, respectively. (b) Retrieved permittivity and permeability inside the complete band gaps close to the frequencies $f_1=1.12$~GHz and $f_2=1.215$~GHz: topological regime is expected when $\varepsilon>0$ and $\mu<0$. Parameter range corresponding to the incomplete bandgap is highlighted by red.}
	\label{fig:4b-extraction}
\end{figure}

Similarly to the previous part, we apply NRW retrieval procedure to the finite sample. However, to obtain consistent results, we choose a boundary of the sample normal to the $\Gamma$K crystallographic direction extracting the  reflection and transmission coefficients at the frequencies of the complete band gaps (Fig.~\ref{fig:4b-extraction}). In turn, the choice of the sample boundary follows from the same requirement of real permittivity and permeability.

Since the band gaps in Fig.~\ref{fig:4b-disp}(a,b) occur at different frequencies, it is quite natural that the metamaterial boundary is chosen differently for these two cases. The choices of the boundary with respect to the unit cell are illustrated in Fig.~\ref{fig:4b-extraction}(a). In this way, we recover the dependence of effective permittivity and permeability on shrinking/expanding parameter $a/R$ presented in Fig.~\ref{fig:4b-extraction}(b).

Inspecting the retrieved material parameters, we observe that the expanded structure with $a/R\leq2.34$ possesses positive permittivity $\eps$ and negative permeability $\mu$ which hints towards its topological nature. On the other hand, shrunken structure with $a/R\geq2.95$ has $\eps<0$ and $\mu>0$ and hence is expected to be trivial.

However, contrary to the case of breathing honeycomb lattice, there is an intermediate range of parameters $2.34<a/R<2.95$ shaded in Fig.~\ref{fig:4b-extraction} by red where the permittivity and permeability are ill-defined. For such geometric parameters, the bandgap in the dispersion is incomplete. As a consequence, depending on the orientation of the sample boundary, an incident wave will be either propagating or evanescent. Therefore, the sign of the effective material parameters retrieved via NRW technique depends on the direction of the wave vector which is a clear fingerprint of spatial dispersion ubiquitous in metamaterials. To avoid  complications related to strong nonlocality, we perform the retrieval only outside of this domain where the results are more consistent with the local effective medium model.

To independently check the topological properties of the modified structure, we evaluate the associated symmetry indices~\cite{Benalcazar2019} which are given by the same expression Eq.~(\ref{eq:symmind}). The results read:
%
\begin{equation}
    \chi^{(6)}_{{\rm e}} = (0,-2)\quad \chi^{(6)}_{{\rm s}} = (0,0),
    \label{eq:ind-4b}
\end{equation}
for expanded ($a/R\leq2.34$) and shrunken lattices ($a/R\geq2.95$), respectively. Note that the topological regimes in a breathing honeycomb lattice and its modified version are inequivalent and correspond to the different primitive generators for $C_6$-symmetric case~\cite{Benalcazar2019}. Nevertheless, our approach based on the behavior of permittivity and permeability captures this type of topological transition as well.



\begin{figure}[t]
	\begin{minipage}{0.9\linewidth}
		\center{\includegraphics[width=1\textwidth]{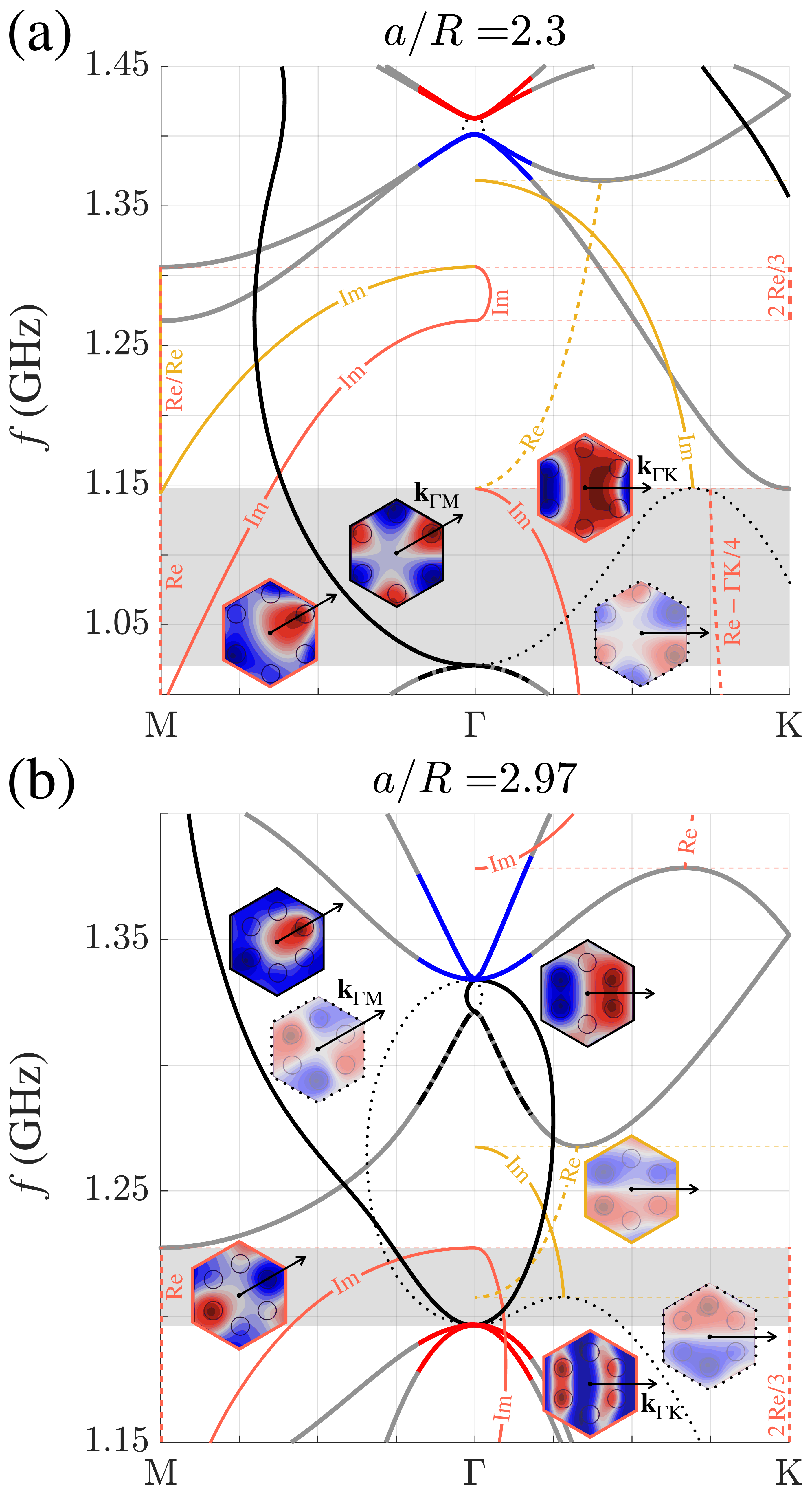}}
	\end{minipage}
	\caption{Dispersion of evanescent modes along M$\Gamma$K path in the vicinity of band gaps for the modified breathing honeycomb structure.  The results for (a) the first band gap near $f_1=1.12$~GHz for the configuration with $a/R=2.3$ and (b) the second gap near $f_2=1.215$~GHz for $a/R=3.25$. Curve designations correspond to Fig.~\ref{fig:3c-disp-weak} caption. Yellow curves describe the complex modes which can not be excited by the wave propagating along $\Gamma$K direction. Solid and dashed lines correspond to the imaginary and real parts of the wave number, respectively. Insets show the typical distribution of $E_z$ component within the unit cell for the different modes.}
	\label{fig:4b-disp-weak}
\end{figure}


At the same time, it should be stressed that the concept of effective medium applied to determine the material parameters is insufficient in many cases. While the effective medium picture assumes that the metamaterial supports a single mode (propagating or evanescent) at a given frequency, the realistic structure supports not only a single dominant wave, but also an infinite set of rapidly decaying evanescent modes. Some of the latter can be excited from air thus contributing to the reflection and transmission coefficients and affecting the results of the retrieval procedure.

To access the physics at in-gap frequencies, we calculate the evanescent modes in the bandgap of the designed metamaterial both for topological and trivial regimes [Fig.~\ref{fig:4b-disp-weak}] using the commercial software package COMSOL Multiphysics, \textit{Weak Form PDE} interface.

The calculated dispersion [Fig.~\ref{fig:4b-disp-weak}] features the same types of evanescent modes with real, imaginary or complex wave numbers as discussed in Sec.~\ref{sec:3C}. However, some of the complex modes can be excited by the incident wave, while other can not. The dispersions of those modes are shown by the orange and yellow curves (less bright in black and white format) in Fig.~\ref{fig:4b-disp-weak}, respectively. 



The obtained results for the complex modes justify our choice of the metamaterial boundary normal to the $\Gamma$K direction. To see that, we examine the expanded structure with $a/R=2.3$ at frequency $f=1.12$~GHz [Fig.~\ref{fig:4b-disp-weak}(a)]. The results suggest that the incident wave propagating along $\Gamma$K excites a single complex mode. On the contrary, if the wave impinges along $\Gamma$M, both complex and imaginary mode with the comparable attenuation lengths are excited. Hence, the effective medium picture is more adequate when the metamaterial is probed from the $\Gamma$K direction.

The situation becomes more involved for the case of a shrunken structure with $a/R=2.97$. Here, both $\Gamma$M and $\Gamma$K directions are associated with the two modes matched by their symmetry to the incident wave. However, for the $\Gamma$K propagation direction and boundary choice shown in Fig.~\ref{fig:4b-extraction}(a) only one complex mode is excited, while the amplitude of the second mode remains small. This in turn enables the conventional retrieval procedure. The calculations presented in  Fig.~\ref{fig:4b-disp-weak} allow us to independently check the results of the retrieval [Fig.~\ref{fig:4b-extraction}(b)] as further discussed in Appendix A.

The observed scenario of a single dominant mode with complex wave number $k'+ik''$ arising in a lossless structure highlights the difficulties of the effective medium treatment. Indeed, assume that such structure is described by the effective permittivity $\eps$ and permeability $\mu$. Since the structure is lossless, both of these quantities should be real. Hence, their product $\eps\,\mu\equiv n^2$ should be real too which is in a clear contradiction with another expression $n^2=(k'+ik'')^2/q^2$, where $q=\omega/c$. Based on that, we conclude that the presence of a single dominant mode with a complex wave number ($k'$ and $k''$ are simultaneously nonzero) is a fingerprint of spatial dispersion effects.

To circumvent this difficulty related to the electromagnetic nonlocality, we exploit the freedom in the choice of the metamaterial boundary. As discussed in Appendix~C, adjusting the position of the boundary, one may tune the phase of the reflection coefficient and thus eliminate the contribution of the real part of the wave number, $k'$. Therefore, we define the effective material parameters as follows:
%
%
\begin{equation}
    \varepsilon\,\mu=n_{\text{eff}}^2=-\left(\frac{c\,k''}{\omega}\right)^2,
    \label{eq:weak-extract}
\end{equation}
where $\omega=2\pi\,f$ is the angular frequency and $k''$ is the imaginary part of the dominant mode wave number. The comparison of the retrieved $\eps\,\mu$ and the calculated right-hand side of Eq.~(\ref{eq:weak-extract}) is provided in Fig.~\ref{fig:apndx-mueps} in Appendix A.

\section{Discussion and conclusions}\label{sec:Discussion}

To conclude, our study bridges a gap between the two views on topological metamaterials: one that exploits the periodic nature of the structure and another relying on the effective medium description. As we prove, the retrieval of effective permittivity and permeability provides a convenient tool to probe the topology of the bands and identify the topological transitions via the change of the material parameters' signs. While the majority of experimental works are investigating the topological states rather than the band topology itself, this technique gives a direct access to the bulk properties. On the other hand, this technique is relatively straightforward from experimental point of view and does not require measurement of the angle-resolved scattering spectra as in ARPES measurements~\cite{LV2019} or its photonic analogs recently applied to study topological metasurfaces~\cite{Gorlach2018}. Furthermore, we anticipate that our observation is valid for the variety of topological structures with the Dirac-type degeneracy near the $\Gamma$ point.


At the same time, the language of effective material parameters suffers from the limitations and inconsistencies. In particular, the boundary conditions at the surface of metamaterial can differ from those expected for the conventional media~\cite{Gorlach2020}, while the existence of multiple evanescent waves or dominant complex modes for the chosen propagation direction strongly affects the results of retrieval even if the metamaterial unit cell is subwavelength. This hints once again towards the essential role of spatial dispersion effects in metamaterials~\cite{Belov2003,Silv2007,Alu2011,Gorlach2015} which can be viewed as local effective media only under very restrictive conditions.

We believe that our study provides interesting insights into topological properties of metamaterials from the effective medium perspective and opens further exciting questions such as effective-medium criteria for higher-order topology.

\section*{Acknowledgments}
Theoretical models were supported by Priority 2030 Federal Academic Leadership Program. Numerical simulations were supported by the Russian Science Foundation (Grant No.~20-72-10065). The authors acknowledge partial support by RPMA grant of School of Physics and Engineering of ITMO University.

\section*{Appendix A. Choice of the metamaterial boundary for the  Nicolson-Ross-Weir retrieval procedure}\label{app:a}

In order to define the metamaterial boundaries properly, we compare the effective material parameters obtained via NRW method for different boundary choices with the results of full-wave numerical simulations of complex band diagrams~\cite{davancco2007complex}. The idea of calculations is to recast the equation for the electric field in such a way that Bloch wavenumber $k$ becomes an eigenvalue in a specified direction at a fixed frequency. The resulting quadratic eigenvalue problem with respect to $k$ is then solved via Weak Form module of Comsol Multiphysics software package.
%
%

\begin{figure}[t]
	\begin{minipage}{0.9\linewidth}
		\center{\includegraphics[width=1\textwidth]{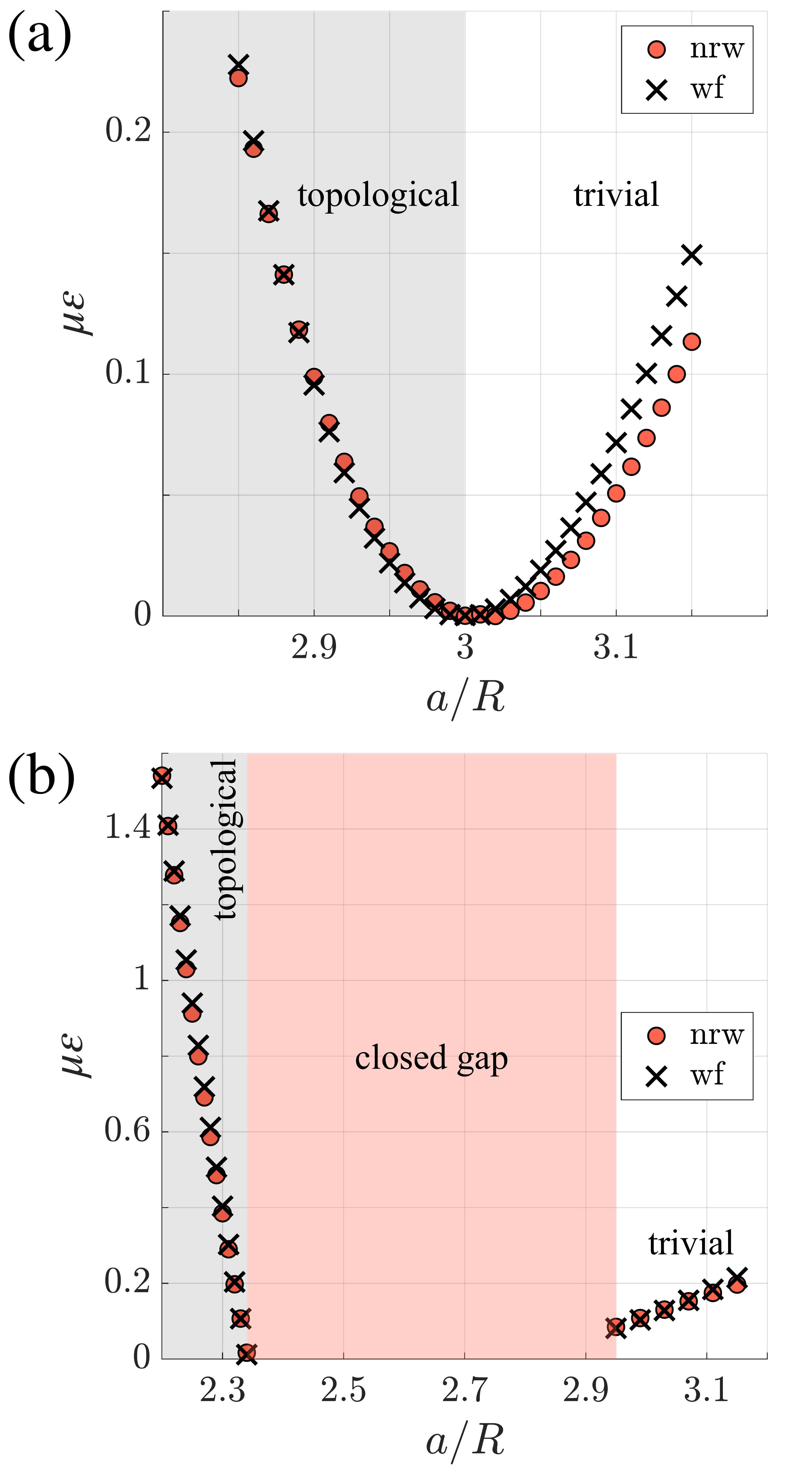}}
	\end{minipage}
	\caption{Squared values of the effective refractive index, $n^2$, for the metamaterials undergoing topological transitions. Panels (a, b) correspond to the conventional and modified breathing honeycomb lattices, respectively. NRW results correspond to the choices of the boundary made in the main text.}
	\label{fig:apndx-mueps}
\end{figure}

The obtained complex band diagrams are shown in Figs.~\ref{fig:3c-disp-weak},\ref{fig:4b-disp-weak}. For each frequency within the band gap, we determine complex wavenumber solutions: $\mathbf{k}=\mathbf{k}'+i\mathbf{k}''$. These solutions correspond to the modes with the field profile $\mathbf{E}(\mathbf{r}) = \mathbf{E_0}e^{i\mathbf{k}\mathbf{r}} = \mathbf{E_0}e^{-\mathbf{k}''\mathbf{r}} e^{i\mathbf{k}'\mathbf{r}}$. Assuming that the contribution of the real part $k'$ can be suppressed via the choice of the boundary, the effective refractive index can be calculated as $n = \sqrt{\eps \mu} = i\,\frac{k''c}{\omega}$. The results of boundary adjustment for both of the considered structures are provided in Fig.~\ref{fig:apndx-mueps}.

Additionally, the calculated complex band diagrams reveal the  direction for the wave vector of impinging wave optimal for NRW retrieval procedure and associated effective medium picture. Specifically, in the case of conventional breathing honeycomb lattice (Fig.~\ref{fig:3c-geom}) the retrieval was performed for $\Gamma$M-direction according to Fig.~\ref{fig:3c-disp-weak}, since in the vicinity of $\Gamma$-point there is only one purely imaginary mode shown by the solid black line. However, $\Gamma$K-direction appears to be more suitable for the modified honeycomb structure (Fig.~\ref{fig:4b-geom}).


\section*{Appendix B. Extracting the signs of $\eps$ and $\mu$}\label{app:b}

NRW method and complex band diagrams allow us to retrieve the absolute values of $\eps$ and $\mu$. To determine their signs, we examine the phase of complex reflection coefficient. 

We assume that the incident field has the structure of the plane wave ${\bf E}({\bf r},t)={\bf E}_0\, e^{i({\bf k}\cdot{\bf r}-\omega t)}$. The relationship between the amplitudes of electric and magnetic field in a medium is given by the impedance $Z=-\frac{i\mu}{\sqrt{|\eps\mu|}}=iZ''$, where  $Z''\equiv\text{Im}(Z)=-\frac{\mu}{\sqrt{|\eps\mu|}}$. The reflection coefficient from the semi-infinite structure is given in turn by the Fresnel formula: $\hat{r}\equiv\frac{E^{\rm{ref}}}{E^{{\rm in}}}=\frac{Z-1}{Z+1}=\frac{iZ''-1}{iZ''+}$, and the dependence of its phase on $Z''$ is illustrated in Fig.~\ref{fig:apndx-imz-geom}.
Based on that plot, we identify the following relations $\mathrm{sign}(\mathrm{arg}(\hat{r})) =- \mathrm{sign}( Z'')=\mathrm{sign}(\mu)=-\mathrm{sign}(\eps)$, where we used the explicit expression for the impedance and took into account different signs of $\eps$ and $\mu$. Thus, inspecting the phase of the reflection coefficient, we can determine which of the material parameters is negative and which is positive. 

\begin{figure}[b]
	\begin{minipage}{0.9\linewidth}
		\center{\includegraphics[width=1\textwidth]{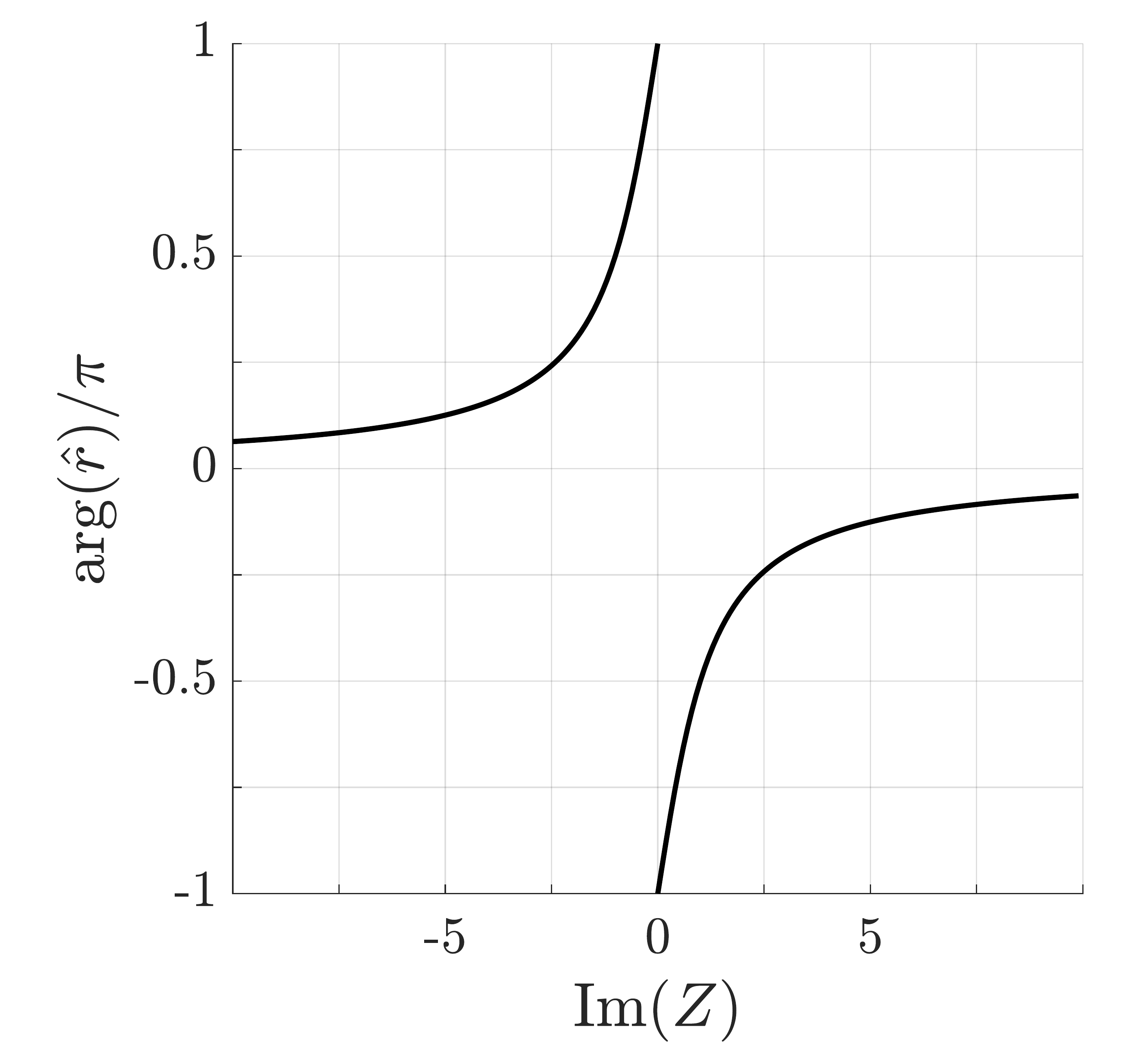}}
	\end{minipage}
	\caption{The phase of the complex reflection coefficient ($\hat{r}$) versus the imaginary part of the impedance, in case  Re~$Z=0$.}
	\label{fig:apndx-imz-geom}
\end{figure}
It should be noted that COMSOL Multiphysics package adopts a different phase convention (${\bf E}({\bf r},t)={\bf E}_0\, e^{-i({\bf k}\cdot{\bf r}-\omega t)}$), due to which the impedance recovered in numerical simulations is complex conjugated.

\section*{Appendix C. Influence of the boundary choice on the retrieval procedure}\label{app:c}

The retrieval of the effective material parameters implies that the realistic metamaterial consisting of the individual inclusions is replaced by the homogeneous slab of the effective medium. Importantly, the boundary of this slab is not uniquely defined. In particular, the shift of the boundary by the value of $h$ directly affects the phase of the reflection coefficient so that $r'=r\,e^{2iq\,h}$. Clearly, any reasonable shifts $h$ of the boundary should be of the order of the lattice constant which is smaller than the wavelength. However, this appears to be sufficient to enable the identification of the effective material parameters, $\eps$ and $\mu$ for the scenario when a single dominant complex mode is present. Specifically, adjusting the position of the boundary, we eliminate the contribution of the real part $k'$ of the wave number.

\begin{figure}[ht]
	\begin{minipage}{0.9\linewidth}
		\center{\includegraphics[width=1\textwidth]{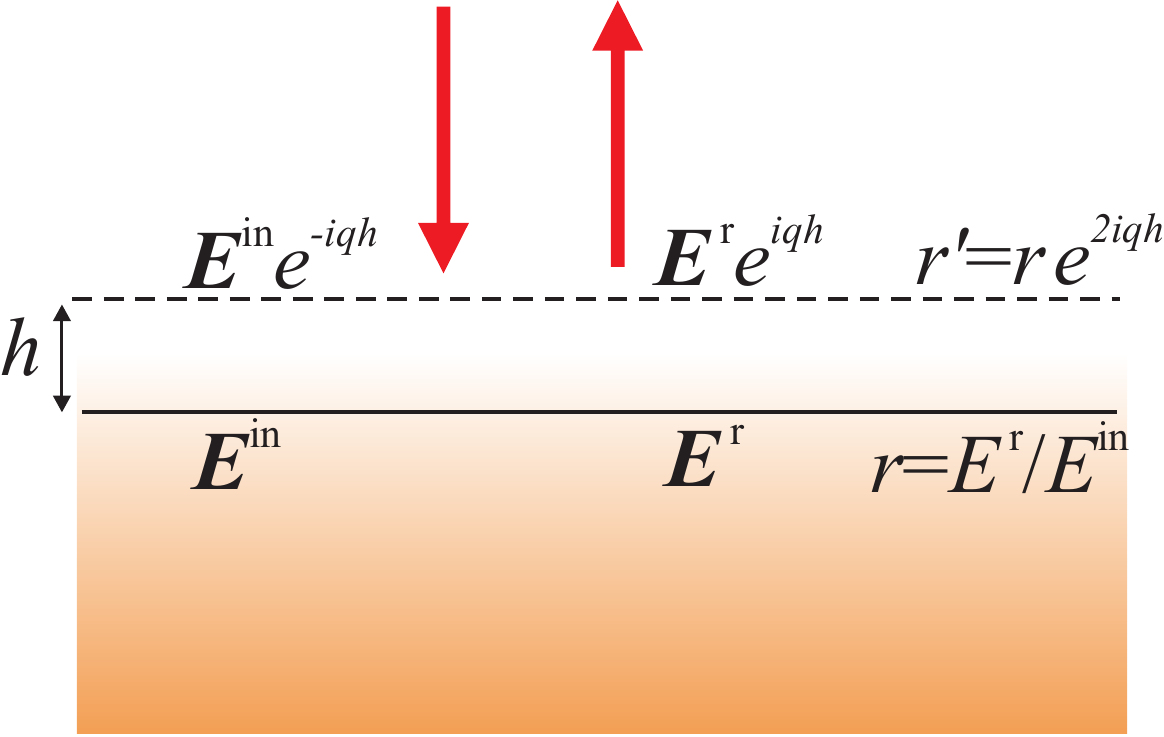}}
	\end{minipage}
	\caption{Dependence of complex reflection coefficient on the choice of the metamaterial boundary. Shift of the boundary by $h$ results in an additional phase of the reflection coefficient equal to $2q\,h$.}
	\label{fig:apndx-imz-geom}
\end{figure}

\newpage
\bibliography{ms}

\begin{thebibliography}{33}%
\makeatletter
\providecommand \@ifxundefined [1]{%
 \@ifx{#1\undefined}
}%
\providecommand \@ifnum [1]{%
 \ifnum #1\expandafter \@firstoftwo
 \else \expandafter \@secondoftwo
 \fi
}%
\providecommand \@ifx [1]{%
 \ifx #1\expandafter \@firstoftwo
 \else \expandafter \@secondoftwo
 \fi
}%
\providecommand \natexlab [1]{#1}%
\providecommand \enquote  [1]{``#1''}%
\providecommand \bibnamefont  [1]{#1}%
\providecommand \bibfnamefont [1]{#1}%
\providecommand \citenamefont [1]{#1}%
\providecommand \href@noop [0]{\@secondoftwo}%
\providecommand \href [0]{\begingroup \@sanitize@url \@href}%
\providecommand \@href[1]{\@@startlink{#1}\@@href}%
\providecommand \@@href[1]{\endgroup#1\@@endlink}%
\providecommand \@sanitize@url [0]{\catcode `\\12\catcode `\$12\catcode
  `\&12\catcode `\#12\catcode `\^12\catcode `\_12\catcode `\%12\relax}%
\providecommand \@@startlink[1]{}%
\providecommand \@@endlink[0]{}%
\providecommand \url  [0]{\begingroup\@sanitize@url \@url }%
\providecommand \@url [1]{\endgroup\@href {#1}{\urlprefix }}%
\providecommand \urlprefix  [0]{URL }%
\providecommand \Eprint [0]{\href }%
\providecommand \doibase [0]{https://doi.org/}%
\providecommand \selectlanguage [0]{\@gobble}%
\providecommand \bibinfo  [0]{\@secondoftwo}%
\providecommand \bibfield  [0]{\@secondoftwo}%
\providecommand \translation [1]{[#1]}%
\providecommand \BibitemOpen [0]{}%
\providecommand \bibitemStop [0]{}%
\providecommand \bibitemNoStop [0]{.\EOS\space}%
\providecommand \EOS [0]{\spacefactor3000\relax}%
\providecommand \BibitemShut  [1]{\csname bibitem#1\endcsname}%
\let\auto@bib@innerbib\@empty
\bibitem [{\citenamefont {Xiao}\ \emph {et~al.}(2010)\citenamefont {Xiao},
  \citenamefont {Chang},\ and\ \citenamefont {Niu}}]{Xiao2010}%
  \BibitemOpen
  \bibfield  {author} {\bibinfo {author} {\bibfnamefont {D.}~\bibnamefont
  {Xiao}}, \bibinfo {author} {\bibfnamefont {M.-C.}\ \bibnamefont {Chang}},\
  and\ \bibinfo {author} {\bibfnamefont {Q.}~\bibnamefont {Niu}},\ }\bibfield
  {title} {\bibinfo {title} {Berry phase effects on electronic properties},\
  }\href {https://doi.org/10.1103/revmodphys.82.1959} {\bibfield  {journal}
  {\bibinfo  {journal} {Reviews of Modern Physics}\ }\textbf {\bibinfo {volume}
  {82}},\ \bibinfo {pages} {1959} (\bibinfo {year} {2010})}\BibitemShut
  {NoStop}%
\bibitem [{\citenamefont {Hasan}\ and\ \citenamefont
  {Kane}(2010)}]{Hasan-Kane}%
  \BibitemOpen
  \bibfield  {author} {\bibinfo {author} {\bibfnamefont {M.~Z.}\ \bibnamefont
  {Hasan}}\ and\ \bibinfo {author} {\bibfnamefont {C.~L.}\ \bibnamefont
  {Kane}},\ }\bibfield  {title} {\bibinfo {title} {Colloquium: Topological
  insulators},\ }\href {https://doi.org/10.1103/RevModPhys.82.3045} {\bibfield
  {journal} {\bibinfo  {journal} {Rev. Mod. Phys.}\ }\textbf {\bibinfo {volume}
  {82}},\ \bibinfo {pages} {3045} (\bibinfo {year} {2010})}\BibitemShut
  {NoStop}%
\bibitem [{\citenamefont {{Lu}}\ \emph {et~al.}(2014)\citenamefont {{Lu}},
  \citenamefont {{Joannopoulos}},\ and\ \citenamefont {{Solja{\v
  c}i{\'c}}}}]{Lu2014}%
  \BibitemOpen
  \bibfield  {author} {\bibinfo {author} {\bibfnamefont {L.}~\bibnamefont
  {{Lu}}}, \bibinfo {author} {\bibfnamefont {J.~D.}\ \bibnamefont
  {{Joannopoulos}}},\ and\ \bibinfo {author} {\bibfnamefont {M.}~\bibnamefont
  {{Solja{\v c}i{\'c}}}},\ }\bibfield  {title} {\bibinfo {title} {{Topological
  photonics}},\ }\href {https://doi.org/10.1038/nphoton.2014.248} {\bibfield
  {journal} {\bibinfo  {journal} {Nat. Photon.}\ }\textbf {\bibinfo {volume}
  {8}},\ \bibinfo {pages} {821} (\bibinfo {year} {2014})}\BibitemShut {NoStop}%
\bibitem [{\citenamefont {Ozawa}\ \emph {et~al.}(2019)\citenamefont {Ozawa},
  \citenamefont {Price}, \citenamefont {Amo}, \citenamefont {Goldman},
  \citenamefont {Hafezi}, \citenamefont {Lu}, \citenamefont {Rechtsman},
  \citenamefont {Schuster}, \citenamefont {Simon}, \citenamefont {Zilberberg},\
  and\ \citenamefont {Carusotto}}]{Ozawa2019}%
  \BibitemOpen
  \bibfield  {author} {\bibinfo {author} {\bibfnamefont {T.}~\bibnamefont
  {Ozawa}}, \bibinfo {author} {\bibfnamefont {H.~M.}\ \bibnamefont {Price}},
  \bibinfo {author} {\bibfnamefont {A.}~\bibnamefont {Amo}}, \bibinfo {author}
  {\bibfnamefont {N.}~\bibnamefont {Goldman}}, \bibinfo {author} {\bibfnamefont
  {M.}~\bibnamefont {Hafezi}}, \bibinfo {author} {\bibfnamefont
  {L.}~\bibnamefont {Lu}}, \bibinfo {author} {\bibfnamefont {M.~C.}\
  \bibnamefont {Rechtsman}}, \bibinfo {author} {\bibfnamefont {D.}~\bibnamefont
  {Schuster}}, \bibinfo {author} {\bibfnamefont {J.}~\bibnamefont {Simon}},
  \bibinfo {author} {\bibfnamefont {O.}~\bibnamefont {Zilberberg}},\ and\
  \bibinfo {author} {\bibfnamefont {I.}~\bibnamefont {Carusotto}},\ }\bibfield
  {title} {\bibinfo {title} {Topological photonics},\ }\href
  {https://doi.org/10.1103/revmodphys.91.015006} {\bibfield  {journal}
  {\bibinfo  {journal} {Reviews of Modern Physics}\ }\textbf {\bibinfo {volume}
  {91}},\ \bibinfo {pages} {015006} (\bibinfo {year} {2019})}\BibitemShut
  {NoStop}%
\bibitem [{\citenamefont {Karzig}\ \emph {et~al.}(2015)\citenamefont {Karzig},
  \citenamefont {Bardyn}, \citenamefont {Lindner},\ and\ \citenamefont
  {Refael}}]{Karzig}%
  \BibitemOpen
  \bibfield  {author} {\bibinfo {author} {\bibfnamefont {T.}~\bibnamefont
  {Karzig}}, \bibinfo {author} {\bibfnamefont {C.-E.}\ \bibnamefont {Bardyn}},
  \bibinfo {author} {\bibfnamefont {N.~H.}\ \bibnamefont {Lindner}},\ and\
  \bibinfo {author} {\bibfnamefont {G.}~\bibnamefont {Refael}},\ }\bibfield
  {title} {\bibinfo {title} {{Topological Polaritons}},\ }\href
  {https://doi.org/10.1103/PhysRevX.5.031001} {\bibfield  {journal} {\bibinfo
  {journal} {Physical Review X}\ }\textbf {\bibinfo {volume} {5}},\ \bibinfo
  {pages} {031001} (\bibinfo {year} {2015})}\BibitemShut {NoStop}%
\bibitem [{\citenamefont {Klembt}\ \emph {et~al.}(2018)\citenamefont {Klembt},
  \citenamefont {Harder}, \citenamefont {Egorov}, \citenamefont {Winkler},
  \citenamefont {Ge}, \citenamefont {Bandres}, \citenamefont {Emmerling},
  \citenamefont {Worschech}, \citenamefont {Liew}, \citenamefont {Segev},
  \citenamefont {Schneider},\ and\ \citenamefont {H{\"o}fling}}]{Klembt2018}%
  \BibitemOpen
  \bibfield  {author} {\bibinfo {author} {\bibfnamefont {S.}~\bibnamefont
  {Klembt}}, \bibinfo {author} {\bibfnamefont {T.~H.}\ \bibnamefont {Harder}},
  \bibinfo {author} {\bibfnamefont {O.~A.}\ \bibnamefont {Egorov}}, \bibinfo
  {author} {\bibfnamefont {K.}~\bibnamefont {Winkler}}, \bibinfo {author}
  {\bibfnamefont {R.}~\bibnamefont {Ge}}, \bibinfo {author} {\bibfnamefont
  {M.~A.}\ \bibnamefont {Bandres}}, \bibinfo {author} {\bibfnamefont
  {M.}~\bibnamefont {Emmerling}}, \bibinfo {author} {\bibfnamefont
  {L.}~\bibnamefont {Worschech}}, \bibinfo {author} {\bibfnamefont {T.~C.~H.}\
  \bibnamefont {Liew}}, \bibinfo {author} {\bibfnamefont {M.}~\bibnamefont
  {Segev}}, \bibinfo {author} {\bibfnamefont {C.}~\bibnamefont {Schneider}},\
  and\ \bibinfo {author} {\bibfnamefont {S.}~\bibnamefont {H{\"o}fling}},\
  }\bibfield  {title} {\bibinfo {title} {Exciton-polariton topological
  insulator},\ }\href {https://doi.org/10.1038/s41586-018-0601-5} {\bibfield
  {journal} {\bibinfo  {journal} {Nature}\ }\textbf {\bibinfo {volume} {562}},\
  \bibinfo {pages} {552} (\bibinfo {year} {2018})}\BibitemShut {NoStop}%
\bibitem [{\citenamefont {Yang}\ \emph {et~al.}(2015)\citenamefont {Yang},
  \citenamefont {Gao}, \citenamefont {Shi}, \citenamefont {Lin}, \citenamefont
  {Gao}, \citenamefont {Chong},\ and\ \citenamefont {Zhang}}]{Yang2015}%
  \BibitemOpen
  \bibfield  {author} {\bibinfo {author} {\bibfnamefont {Z.}~\bibnamefont
  {Yang}}, \bibinfo {author} {\bibfnamefont {F.}~\bibnamefont {Gao}}, \bibinfo
  {author} {\bibfnamefont {X.}~\bibnamefont {Shi}}, \bibinfo {author}
  {\bibfnamefont {X.}~\bibnamefont {Lin}}, \bibinfo {author} {\bibfnamefont
  {Z.}~\bibnamefont {Gao}}, \bibinfo {author} {\bibfnamefont {Y.}~\bibnamefont
  {Chong}},\ and\ \bibinfo {author} {\bibfnamefont {B.}~\bibnamefont {Zhang}},\
  }\bibfield  {title} {\bibinfo {title} {{Topological Acoustics}},\ }\href
  {https://doi.org/10.1103/PhysRevLett.114.114301} {\bibfield  {journal}
  {\bibinfo  {journal} {Phys. Rev. Lett.}\ }\textbf {\bibinfo {volume} {114}},\
  \bibinfo {pages} {114301} (\bibinfo {year} {2015})}\BibitemShut {NoStop}%
\bibitem [{\citenamefont {Huber}(2016)}]{Huber2016}%
  \BibitemOpen
  \bibfield  {author} {\bibinfo {author} {\bibfnamefont {S.~D.}\ \bibnamefont
  {Huber}},\ }\bibfield  {title} {\bibinfo {title} {Topological mechanics},\
  }\href {https://doi.org/10.1038/nphys3801} {\bibfield  {journal} {\bibinfo
  {journal} {Nature Physics}\ }\textbf {\bibinfo {volume} {12}},\ \bibinfo
  {pages} {621} (\bibinfo {year} {2016})}\BibitemShut {NoStop}%
\bibitem [{\citenamefont {Benalcazar}\ \emph {et~al.}(2019)\citenamefont
  {Benalcazar}, \citenamefont {Li},\ and\ \citenamefont
  {Hughes}}]{Benalcazar2019}%
  \BibitemOpen
  \bibfield  {author} {\bibinfo {author} {\bibfnamefont {W.~A.}\ \bibnamefont
  {Benalcazar}}, \bibinfo {author} {\bibfnamefont {T.}~\bibnamefont {Li}},\
  and\ \bibinfo {author} {\bibfnamefont {T.~L.}\ \bibnamefont {Hughes}},\
  }\bibfield  {title} {\bibinfo {title} {{Quantization of fractional corner
  charge in $C_n$-symmetric higher-order topological crystalline insulators}},\
  }\href {https://doi.org/10.1103/PhysRevB.99.245151} {\bibfield  {journal}
  {\bibinfo  {journal} {Physical Review B}\ }\textbf {\bibinfo {volume} {99}},\
  \bibinfo {pages} {245151} (\bibinfo {year} {2019})}\BibitemShut {NoStop}%
\bibitem [{\citenamefont {Silveirinha}(2015)}]{Silv2015}%
  \BibitemOpen
  \bibfield  {author} {\bibinfo {author} {\bibfnamefont {M.~G.}\ \bibnamefont
  {Silveirinha}},\ }\bibfield  {title} {\bibinfo {title} {Chern invariants for
  continuous media},\ }\href {https://doi.org/10.1103/physrevb.92.125153}
  {\bibfield  {journal} {\bibinfo  {journal} {Physical Review B}\ }\textbf
  {\bibinfo {volume} {92}},\ \bibinfo {pages} {125153} (\bibinfo {year}
  {2015})}\BibitemShut {NoStop}%
\bibitem [{\citenamefont {Silveirinha}(2016{\natexlab{a}})}]{Silveirinha2016}%
  \BibitemOpen
  \bibfield  {author} {\bibinfo {author} {\bibfnamefont {M.~G.}\ \bibnamefont
  {Silveirinha}},\ }\bibfield  {title} {\bibinfo {title} {{$Z_2$ topological
  index for continuous photonic materials}},\ }\href
  {https://doi.org/10.1103/physrevb.93.075110} {\bibfield  {journal} {\bibinfo
  {journal} {Physical Review B}\ }\textbf {\bibinfo {volume} {93}},\ \bibinfo
  {pages} {075110} (\bibinfo {year} {2016}{\natexlab{a}})}\BibitemShut
  {NoStop}%
\bibitem [{\citenamefont {Silveirinha}(2016{\natexlab{b}})}]{Silv2016}%
  \BibitemOpen
  \bibfield  {author} {\bibinfo {author} {\bibfnamefont {M.~G.}\ \bibnamefont
  {Silveirinha}},\ }\bibfield  {title} {\bibinfo {title} {Bulk-edge
  correspondence for topological photonic continua},\ }\href
  {https://doi.org/10.1103/physrevb.94.205105} {\bibfield  {journal} {\bibinfo
  {journal} {Physical Review B}\ }\textbf {\bibinfo {volume} {94}},\ \bibinfo
  {pages} {205105} (\bibinfo {year} {2016}{\natexlab{b}})}\BibitemShut
  {NoStop}%
\bibitem [{\citenamefont {Pakniyat}\ \emph {et~al.}(2022)\citenamefont
  {Pakniyat}, \citenamefont {Gangaraj},\ and\ \citenamefont
  {Hanson}}]{Hanson2022}%
  \BibitemOpen
  \bibfield  {author} {\bibinfo {author} {\bibfnamefont {S.}~\bibnamefont
  {Pakniyat}}, \bibinfo {author} {\bibfnamefont {S.~A.~H.}\ \bibnamefont
  {Gangaraj}},\ and\ \bibinfo {author} {\bibfnamefont {G.~W.}\ \bibnamefont
  {Hanson}},\ }\bibfield  {title} {\bibinfo {title} {{Chern invariants of
  topological continua: A self-consistent nonlocal hydrodynamic model}},\
  }\href {https://doi.org/10.1103/physrevb.105.035310} {\bibfield  {journal}
  {\bibinfo  {journal} {Physical Review B}\ }\textbf {\bibinfo {volume}
  {105}},\ \bibinfo {pages} {035310} (\bibinfo {year} {2022})}\BibitemShut
  {NoStop}%
\bibitem [{\citenamefont {Wu}\ and\ \citenamefont {Hu}(2015)}]{Wu}%
  \BibitemOpen
  \bibfield  {author} {\bibinfo {author} {\bibfnamefont {L.-H.}\ \bibnamefont
  {Wu}}\ and\ \bibinfo {author} {\bibfnamefont {X.}~\bibnamefont {Hu}},\
  }\bibfield  {title} {\bibinfo {title} {{Scheme for Achieving a Topological
  Photonic Crystal by Using Dielectric Material}},\ }\href
  {https://doi.org/10.1103/physrevlett.114.223901} {\bibfield  {journal}
  {\bibinfo  {journal} {Physical Review Letters}\ }\textbf {\bibinfo {volume}
  {114}},\ \bibinfo {pages} {223901} (\bibinfo {year} {2015})}\BibitemShut
  {NoStop}%
\bibitem [{\citenamefont {Yves}\ \emph {et~al.}(2017)\citenamefont {Yves},
  \citenamefont {Fleury}, \citenamefont {Berthelot}, \citenamefont {Fink},
  \citenamefont {Lemoult},\ and\ \citenamefont {Lerosey}}]{Yves2017}%
  \BibitemOpen
  \bibfield  {author} {\bibinfo {author} {\bibfnamefont {S.}~\bibnamefont
  {Yves}}, \bibinfo {author} {\bibfnamefont {R.}~\bibnamefont {Fleury}},
  \bibinfo {author} {\bibfnamefont {T.}~\bibnamefont {Berthelot}}, \bibinfo
  {author} {\bibfnamefont {M.}~\bibnamefont {Fink}}, \bibinfo {author}
  {\bibfnamefont {F.}~\bibnamefont {Lemoult}},\ and\ \bibinfo {author}
  {\bibfnamefont {G.}~\bibnamefont {Lerosey}},\ }\bibfield  {title} {\bibinfo
  {title} {Crystalline metamaterials for topological properties at
  subwavelength scales},\ }\href {https://doi.org/10.1038/ncomms16023}
  {\bibfield  {journal} {\bibinfo  {journal} {Nature Communications}\ }\textbf
  {\bibinfo {volume} {8}},\ \bibinfo {pages} {16023} (\bibinfo {year}
  {2017})}\BibitemShut {NoStop}%
\bibitem [{\citenamefont {Li}\ \emph {et~al.}(2018)\citenamefont {Li},
  \citenamefont {Sun}, \citenamefont {Zhu}, \citenamefont {Guo}, \citenamefont
  {Jiang}, \citenamefont {Kariyado}, \citenamefont {Chen},\ and\ \citenamefont
  {Hu}}]{Li2018}%
  \BibitemOpen
  \bibfield  {author} {\bibinfo {author} {\bibfnamefont {Y.}~\bibnamefont
  {Li}}, \bibinfo {author} {\bibfnamefont {Y.}~\bibnamefont {Sun}}, \bibinfo
  {author} {\bibfnamefont {W.}~\bibnamefont {Zhu}}, \bibinfo {author}
  {\bibfnamefont {Z.}~\bibnamefont {Guo}}, \bibinfo {author} {\bibfnamefont
  {J.}~\bibnamefont {Jiang}}, \bibinfo {author} {\bibfnamefont
  {T.}~\bibnamefont {Kariyado}}, \bibinfo {author} {\bibfnamefont
  {H.}~\bibnamefont {Chen}},\ and\ \bibinfo {author} {\bibfnamefont
  {X.}~\bibnamefont {Hu}},\ }\bibfield  {title} {\bibinfo {title} {Topological
  {LC}-circuits based on microstrips and observation of electromagnetic modes
  with orbital angular momentum},\ }\href
  {https://doi.org/10.1038/s41467-018-07084-2} {\bibfield  {journal} {\bibinfo
  {journal} {Nature Communications}\ }\textbf {\bibinfo {volume} {9}},\
  \bibinfo {pages} {4598} (\bibinfo {year} {2018})}\BibitemShut {NoStop}%
\bibitem [{\citenamefont {Yang}\ \emph {et~al.}(2018)\citenamefont {Yang},
  \citenamefont {Xu}, \citenamefont {Xu}, \citenamefont {Wang}, \citenamefont
  {Jiang}, \citenamefont {Hu},\ and\ \citenamefont {Hang}}]{Yang2018}%
  \BibitemOpen
  \bibfield  {author} {\bibinfo {author} {\bibfnamefont {Y.}~\bibnamefont
  {Yang}}, \bibinfo {author} {\bibfnamefont {Y.~F.}\ \bibnamefont {Xu}},
  \bibinfo {author} {\bibfnamefont {T.}~\bibnamefont {Xu}}, \bibinfo {author}
  {\bibfnamefont {H.-X.}\ \bibnamefont {Wang}}, \bibinfo {author}
  {\bibfnamefont {J.-H.}\ \bibnamefont {Jiang}}, \bibinfo {author}
  {\bibfnamefont {X.}~\bibnamefont {Hu}},\ and\ \bibinfo {author}
  {\bibfnamefont {Z.}~\bibnamefont {Hang}},\ }\bibfield  {title} {\bibinfo
  {title} {{Visualization of a Unidirectional Electromagnetic Waveguide Using
  Topological Photonic Crystals Made of Dielectric Materials}},\ }\href
  {https://doi.org/10.1103/physrevlett.120.217401} {\bibfield  {journal}
  {\bibinfo  {journal} {Physical Review Letters}\ }\textbf {\bibinfo {volume}
  {120}},\ \bibinfo {pages} {217401} (\bibinfo {year} {2018})}\BibitemShut
  {NoStop}%
\bibitem [{\citenamefont {Barik}\ \emph {et~al.}(2018)\citenamefont {Barik},
  \citenamefont {Karasahin}, \citenamefont {Flower}, \citenamefont {Cai},
  \citenamefont {Miyake}, \citenamefont {DeGottardi}, \citenamefont {Hafezi},\
  and\ \citenamefont {Waks}}]{Barik2018}%
  \BibitemOpen
  \bibfield  {author} {\bibinfo {author} {\bibfnamefont {S.}~\bibnamefont
  {Barik}}, \bibinfo {author} {\bibfnamefont {A.}~\bibnamefont {Karasahin}},
  \bibinfo {author} {\bibfnamefont {C.}~\bibnamefont {Flower}}, \bibinfo
  {author} {\bibfnamefont {T.}~\bibnamefont {Cai}}, \bibinfo {author}
  {\bibfnamefont {H.}~\bibnamefont {Miyake}}, \bibinfo {author} {\bibfnamefont
  {W.}~\bibnamefont {DeGottardi}}, \bibinfo {author} {\bibfnamefont
  {M.}~\bibnamefont {Hafezi}},\ and\ \bibinfo {author} {\bibfnamefont
  {E.}~\bibnamefont {Waks}},\ }\bibfield  {title} {\bibinfo {title} {A
  topological quantum optics interface},\ }\href
  {https://doi.org/10.1126/science.aaq0327} {\bibfield  {journal} {\bibinfo
  {journal} {Science}\ }\textbf {\bibinfo {volume} {359}},\ \bibinfo {pages}
  {666} (\bibinfo {year} {2018})}\BibitemShut {NoStop}%
\bibitem [{\citenamefont {Noh}\ \emph {et~al.}(2018)\citenamefont {Noh},
  \citenamefont {Benalcazar}, \citenamefont {Huang}, \citenamefont {Collins},
  \citenamefont {Chen}, \citenamefont {Hughes},\ and\ \citenamefont
  {Rechtsman}}]{Noh2018}%
  \BibitemOpen
  \bibfield  {author} {\bibinfo {author} {\bibfnamefont {J.}~\bibnamefont
  {Noh}}, \bibinfo {author} {\bibfnamefont {W.~A.}\ \bibnamefont {Benalcazar}},
  \bibinfo {author} {\bibfnamefont {S.}~\bibnamefont {Huang}}, \bibinfo
  {author} {\bibfnamefont {M.~J.}\ \bibnamefont {Collins}}, \bibinfo {author}
  {\bibfnamefont {K.~P.}\ \bibnamefont {Chen}}, \bibinfo {author}
  {\bibfnamefont {T.~L.}\ \bibnamefont {Hughes}},\ and\ \bibinfo {author}
  {\bibfnamefont {M.~C.}\ \bibnamefont {Rechtsman}},\ }\bibfield  {title}
  {\bibinfo {title} {Topological protection of photonic mid-gap defect modes},\
  }\href {https://doi.org/10.1038/s41566-018-0179-3} {\bibfield  {journal}
  {\bibinfo  {journal} {Nature Photonics}\ }\textbf {\bibinfo {volume} {12}},\
  \bibinfo {pages} {408} (\bibinfo {year} {2018})}\BibitemShut {NoStop}%
\bibitem [{\citenamefont {Gorlach}\ \emph {et~al.}(2018)\citenamefont
  {Gorlach}, \citenamefont {Ni}, \citenamefont {Smirnova}, \citenamefont
  {Korobkin}, \citenamefont {Zhirihin}, \citenamefont {Slobozhanyuk},
  \citenamefont {Belov}, \citenamefont {Al{\`{u}}},\ and\ \citenamefont
  {Khanikaev}}]{Gorlach2018}%
  \BibitemOpen
  \bibfield  {author} {\bibinfo {author} {\bibfnamefont {M.~A.}\ \bibnamefont
  {Gorlach}}, \bibinfo {author} {\bibfnamefont {X.}~\bibnamefont {Ni}},
  \bibinfo {author} {\bibfnamefont {D.~A.}\ \bibnamefont {Smirnova}}, \bibinfo
  {author} {\bibfnamefont {D.}~\bibnamefont {Korobkin}}, \bibinfo {author}
  {\bibfnamefont {D.}~\bibnamefont {Zhirihin}}, \bibinfo {author}
  {\bibfnamefont {A.~P.}\ \bibnamefont {Slobozhanyuk}}, \bibinfo {author}
  {\bibfnamefont {P.~A.}\ \bibnamefont {Belov}}, \bibinfo {author}
  {\bibfnamefont {A.}~\bibnamefont {Al{\`{u}}}},\ and\ \bibinfo {author}
  {\bibfnamefont {A.~B.}\ \bibnamefont {Khanikaev}},\ }\bibfield  {title}
  {\bibinfo {title} {Far-field probing of leaky topological states in
  all-dielectric metasurfaces},\ }\href
  {https://doi.org/10.1038/s41467-018-03330-9} {\bibfield  {journal} {\bibinfo
  {journal} {Nature Communications}\ }\textbf {\bibinfo {volume} {9}},\
  \bibinfo {pages} {909} (\bibinfo {year} {2018})}\BibitemShut {NoStop}%
\bibitem [{\citenamefont {Smirnova}\ \emph {et~al.}(2019)\citenamefont
  {Smirnova}, \citenamefont {Kruk}, \citenamefont {Leykam}, \citenamefont
  {Melik-Gaykazyan}, \citenamefont {Choi},\ and\ \citenamefont
  {Kivshar}}]{Smirnova2019}%
  \BibitemOpen
  \bibfield  {author} {\bibinfo {author} {\bibfnamefont {D.}~\bibnamefont
  {Smirnova}}, \bibinfo {author} {\bibfnamefont {S.}~\bibnamefont {Kruk}},
  \bibinfo {author} {\bibfnamefont {D.}~\bibnamefont {Leykam}}, \bibinfo
  {author} {\bibfnamefont {E.}~\bibnamefont {Melik-Gaykazyan}}, \bibinfo
  {author} {\bibfnamefont {D.-Y.}\ \bibnamefont {Choi}},\ and\ \bibinfo
  {author} {\bibfnamefont {Y.}~\bibnamefont {Kivshar}},\ }\bibfield  {title}
  {\bibinfo {title} {Third-harmonic generation in photonic topological
  metasurfaces},\ }\href {https://doi.org/10.1103/physrevlett.123.103901}
  {\bibfield  {journal} {\bibinfo  {journal} {Physical Review Letters}\
  }\textbf {\bibinfo {volume} {123}},\ \bibinfo {pages} {103901} (\bibinfo
  {year} {2019})}\BibitemShut {NoStop}%
\bibitem [{\citenamefont {Parappurath}\ \emph {et~al.}(2020)\citenamefont
  {Parappurath}, \citenamefont {Alpeggiani}, \citenamefont {Kuipers},\ and\
  \citenamefont {Verhagen}}]{Kuipers2020}%
  \BibitemOpen
  \bibfield  {author} {\bibinfo {author} {\bibfnamefont {N.}~\bibnamefont
  {Parappurath}}, \bibinfo {author} {\bibfnamefont {F.}~\bibnamefont
  {Alpeggiani}}, \bibinfo {author} {\bibfnamefont {L.}~\bibnamefont
  {Kuipers}},\ and\ \bibinfo {author} {\bibfnamefont {E.}~\bibnamefont
  {Verhagen}},\ }\bibfield  {title} {\bibinfo {title} {{Direct observation of
  topological edge states in silicon photonic crystals: Spin, dispersion, and
  chiral routing}},\ }\href {https://doi.org/10.1126/sciadv.aaw4137} {\bibfield
   {journal} {\bibinfo  {journal} {Science Advances}\ }\textbf {\bibinfo
  {volume} {6}},\ \bibinfo {pages} {eaaw4137} (\bibinfo {year}
  {2020})}\BibitemShut {NoStop}%
\bibitem [{\citenamefont {Nicolson}\ and\ \citenamefont
  {Ross}(1970)}]{Nicolson}%
  \BibitemOpen
  \bibfield  {author} {\bibinfo {author} {\bibfnamefont {A.~M.}\ \bibnamefont
  {Nicolson}}\ and\ \bibinfo {author} {\bibfnamefont {G.~F.}\ \bibnamefont
  {Ross}},\ }\bibfield  {title} {\bibinfo {title} {{Measurement of the
  Intrinsic Properties of Materials by Time-Domain Techniques}},\ }\href
  {https://doi.org/10.1109/tim.1970.4313932} {\bibfield  {journal} {\bibinfo
  {journal} {{IEEE Transactions on Instrumentation and Measurement}}\ }\textbf
  {\bibinfo {volume} {19}},\ \bibinfo {pages} {377} (\bibinfo {year}
  {1970})}\BibitemShut {NoStop}%
\bibitem [{\citenamefont {Weir}(1974)}]{Weir}%
  \BibitemOpen
  \bibfield  {author} {\bibinfo {author} {\bibfnamefont {W.}~\bibnamefont
  {Weir}},\ }\bibfield  {title} {\bibinfo {title} {{Automatic Measurement of
  Complex Dielectric Constant and Permeability and Microwave Frequencies}},\
  }\href@noop {} {\bibfield  {journal} {\bibinfo  {journal} {Proceedings of the
  IEEE}\ }\textbf {\bibinfo {volume} {62}},\ \bibinfo {pages} {33} (\bibinfo
  {year} {1974})}\BibitemShut {NoStop}%
\bibitem [{\citenamefont {Davan{\c{c}}o}\ \emph {et~al.}(2007)\citenamefont
  {Davan{\c{c}}o}, \citenamefont {Urzhumov},\ and\ \citenamefont
  {Shvets}}]{davancco2007complex}%
  \BibitemOpen
  \bibfield  {author} {\bibinfo {author} {\bibfnamefont {M.}~\bibnamefont
  {Davan{\c{c}}o}}, \bibinfo {author} {\bibfnamefont {Y.}~\bibnamefont
  {Urzhumov}},\ and\ \bibinfo {author} {\bibfnamefont {G.}~\bibnamefont
  {Shvets}},\ }\bibfield  {title} {\bibinfo {title} {The complex bloch bands of
  a 2d plasmonic crystal displaying isotropic negative refraction},\ }\href
  {https://doi.org/10.1364/OE.15.009681} {\bibfield  {journal} {\bibinfo
  {journal} {Optics Express}\ }\textbf {\bibinfo {volume} {15}},\ \bibinfo
  {pages} {9681} (\bibinfo {year} {2007})}\BibitemShut {NoStop}%
\bibitem [{\citenamefont {Fietz}\ \emph {et~al.}(2011)\citenamefont {Fietz},
  \citenamefont {Urzhumov},\ and\ \citenamefont {Shvets}}]{fietz2011complex}%
  \BibitemOpen
  \bibfield  {author} {\bibinfo {author} {\bibfnamefont {C.}~\bibnamefont
  {Fietz}}, \bibinfo {author} {\bibfnamefont {Y.}~\bibnamefont {Urzhumov}},\
  and\ \bibinfo {author} {\bibfnamefont {G.}~\bibnamefont {Shvets}},\
  }\bibfield  {title} {\bibinfo {title} {{Complex k band diagrams of 3D
  metamaterial/photonic crystals}},\ }\href
  {https://doi.org/10.1364/OE.19.019027} {\bibfield  {journal} {\bibinfo
  {journal} {Optics Express}\ }\textbf {\bibinfo {volume} {19}},\ \bibinfo
  {pages} {19027} (\bibinfo {year} {2011})}\BibitemShut {NoStop}%
\bibitem [{\citenamefont {Gorlach}\ and\ \citenamefont
  {Lapine}(2020)}]{Gorlach2020}%
  \BibitemOpen
  \bibfield  {author} {\bibinfo {author} {\bibfnamefont {M.~A.}\ \bibnamefont
  {Gorlach}}\ and\ \bibinfo {author} {\bibfnamefont {M.}~\bibnamefont
  {Lapine}},\ }\bibfield  {title} {\bibinfo {title} {Boundary conditions for
  the effective-medium description of subwavelength multilayered structures},\
  }\href {https://doi.org/10.1103/physrevb.101.075127} {\bibfield  {journal}
  {\bibinfo  {journal} {Physical Review B}\ }\textbf {\bibinfo {volume}
  {101}},\ \bibinfo {pages} {075127} (\bibinfo {year} {2020})}\BibitemShut
  {NoStop}%
\bibitem [{\citenamefont {Wu}\ \emph {et~al.}(2021)\citenamefont {Wu},
  \citenamefont {Jiang}, \citenamefont {Liu},\ and\ \citenamefont
  {Jiang}}]{Wu2021}%
  \BibitemOpen
  \bibfield  {author} {\bibinfo {author} {\bibfnamefont {S.}~\bibnamefont
  {Wu}}, \bibinfo {author} {\bibfnamefont {B.}~\bibnamefont {Jiang}}, \bibinfo
  {author} {\bibfnamefont {Y.}~\bibnamefont {Liu}},\ and\ \bibinfo {author}
  {\bibfnamefont {J.-H.}\ \bibnamefont {Jiang}},\ }\bibfield  {title} {\bibinfo
  {title} {All-dielectric photonic crystal with unconventional higher-order
  topology},\ }\href {https://doi.org/10.1364/prj.418689} {\bibfield  {journal}
  {\bibinfo  {journal} {Photonics Research}\ }\textbf {\bibinfo {volume} {9}},\
  \bibinfo {pages} {668} (\bibinfo {year} {2021})}\BibitemShut {NoStop}%
\bibitem [{\citenamefont {Lv}\ \emph {et~al.}(2019)\citenamefont {Lv},
  \citenamefont {Qian},\ and\ \citenamefont {Ding}}]{LV2019}%
  \BibitemOpen
  \bibfield  {author} {\bibinfo {author} {\bibfnamefont {B.}~\bibnamefont
  {Lv}}, \bibinfo {author} {\bibfnamefont {T.}~\bibnamefont {Qian}},\ and\
  \bibinfo {author} {\bibfnamefont {H.}~\bibnamefont {Ding}},\ }\bibfield
  {title} {\bibinfo {title} {Angle-resolved photoemission spectroscopy and its
  application to topological materials},\ }\href
  {https://doi.org/10.1038/s42254-019-0088-5} {\bibfield  {journal} {\bibinfo
  {journal} {Nature Reviews Physics}\ }\textbf {\bibinfo {volume} {1}},\
  \bibinfo {pages} {609} (\bibinfo {year} {2019})}\BibitemShut {NoStop}%
\bibitem [{\citenamefont {Belov}\ \emph {et~al.}(2003)\citenamefont {Belov},
  \citenamefont {Marqu{\'{e}}s}, \citenamefont {Maslovski}, \citenamefont
  {Nefedov}, \citenamefont {Silveirinha}, \citenamefont {Simovski},\ and\
  \citenamefont {Tretyakov}}]{Belov2003}%
  \BibitemOpen
  \bibfield  {author} {\bibinfo {author} {\bibfnamefont {P.~A.}\ \bibnamefont
  {Belov}}, \bibinfo {author} {\bibfnamefont {R.}~\bibnamefont
  {Marqu{\'{e}}s}}, \bibinfo {author} {\bibfnamefont {S.~I.}\ \bibnamefont
  {Maslovski}}, \bibinfo {author} {\bibfnamefont {I.~S.}\ \bibnamefont
  {Nefedov}}, \bibinfo {author} {\bibfnamefont {M.}~\bibnamefont
  {Silveirinha}}, \bibinfo {author} {\bibfnamefont {C.~R.}\ \bibnamefont
  {Simovski}},\ and\ \bibinfo {author} {\bibfnamefont {S.~A.}\ \bibnamefont
  {Tretyakov}},\ }\bibfield  {title} {\bibinfo {title} {Strong spatial
  dispersion in wire media in the very large wavelength limit},\ }\href
  {https://doi.org/10.1103/physrevb.67.113103} {\bibfield  {journal} {\bibinfo
  {journal} {Physical Review B}\ }\textbf {\bibinfo {volume} {67}},\ \bibinfo
  {pages} {113103} (\bibinfo {year} {2003})}\BibitemShut {NoStop}%
\bibitem [{\citenamefont {Silveirinha}(2007)}]{Silv2007}%
  \BibitemOpen
  \bibfield  {author} {\bibinfo {author} {\bibfnamefont {M.~G.}\ \bibnamefont
  {Silveirinha}},\ }\bibfield  {title} {\bibinfo {title} {Metamaterial
  homogenization approach with application to the characterization of
  microstructured composites with negative parameters},\ }\href
  {https://doi.org/10.1103/physrevb.75.115104} {\bibfield  {journal} {\bibinfo
  {journal} {Physical Review B}\ }\textbf {\bibinfo {volume} {75}},\ \bibinfo
  {pages} {115104} (\bibinfo {year} {2007})}\BibitemShut {NoStop}%
\bibitem [{\citenamefont {Al{\`{u}}}(2011)}]{Alu2011}%
  \BibitemOpen
  \bibfield  {author} {\bibinfo {author} {\bibfnamefont {A.}~\bibnamefont
  {Al{\`{u}}}},\ }\bibfield  {title} {\bibinfo {title} {First-principles
  homogenization theory for periodic metamaterials},\ }\href
  {https://doi.org/10.1103/physrevb.84.075153} {\bibfield  {journal} {\bibinfo
  {journal} {Physical Review B}\ }\textbf {\bibinfo {volume} {84}},\ \bibinfo
  {pages} {075153} (\bibinfo {year} {2011})}\BibitemShut {NoStop}%
\bibitem [{\citenamefont {Gorlach}\ and\ \citenamefont
  {Belov}(2015)}]{Gorlach2015}%
  \BibitemOpen
  \bibfield  {author} {\bibinfo {author} {\bibfnamefont {M.~A.}\ \bibnamefont
  {Gorlach}}\ and\ \bibinfo {author} {\bibfnamefont {P.~A.}\ \bibnamefont
  {Belov}},\ }\bibfield  {title} {\bibinfo {title} {Nonlocality in uniaxially
  polarizable media},\ }\href {https://doi.org/10.1103/physrevb.92.085107}
  {\bibfield  {journal} {\bibinfo  {journal} {Physical Review B}\ }\textbf
  {\bibinfo {volume} {92}},\ \bibinfo {pages} {085107} (\bibinfo {year}
  {2015})}\BibitemShut {NoStop}%
\end{thebibliography}%

\end{document}